\documentclass{aastex63}
\usepackage{amsmath}
\newcommand{\be}{\begin{eqnarray}}
\newcommand{\ee}{\end{eqnarray}}
\def\la{\mathbin{\lower 3pt\hbox
      {$\rlap{\raise 5pt\hbox{$\char'074$}}\mathchar"7218$}}}
\def\ga{\mathbin{\lower 3pt\hbox
      {$\rlap{\raise 5pt\hbox{$\char'076$}}\mathchar"7218$}}}
\renewcommand{\vec}[1]{\mathbf{#1}}
\newcommand{\grad}{\mathbf{\nabla}}
\newcommand\lya{Ly$\alpha$\ }

\shorttitle{Resonant Scattering}
\shortauthors{McClellan et al.}

\graphicspath{./}

\begin{document}

\title{A Novel Solution for Resonant Scattering Using Self-Consistent Boundary Conditions}

\correspondingauthor{B. Connor McClellan}
\email{bcm2vn@virginia.edu}

\author[0000-0002-6040-8281]{B. Connor McClellan}
\author[0000-0001-7488-4468]{Shane Davis}
\author[0000-0001-5611-1349]{Phil Arras}
\affiliation{Department of Astronomy, University of Virginia, Charlottesville, VA 22904, USA}

\begin{abstract}

We present two novel additions to the semi-analytic solution of Lyman $\alpha$ (Ly$\alpha$) radiative transfer in spherical geometry: (1) implementation of the correct boundary condition for a steady source, and (2) solution of the time-dependent problem for an impulsive source. For the steady-state problem, the solution can be represented as a sum of two terms: a previously-known analytic solution of the equation with  mean intensity $J=0$ at the surface, and a novel, semi-analytic solution which enforces the correct boundary condition of zero-ingoing intensity at the surface. This solution is compared to that of the Monte Carlo method, which is valid at arbitrary optical depth. It is shown that the size of the correction is of order unity when the spectral peaks approach the Doppler core and decreases slowly with line center optical depth, specifically as $(a \tau_0)^{-1/3}$, which may explain discrepancies seen in previous studies. For the impulsive problem, the time, spatial, and frequency dependence of the solution are expressed using an eigenfunction expansion in order to characterize the escape time distribution and emergent spectra of photons. It is shown that the lowest-order eigenfrequency agrees well with the decay rate found in the Monte Carlo escape time distribution at sufficiently large line-center optical depths. The characterization of the escape-time distribution highlights the potential for a Monte Carlo acceleration method, which would sample photon escape properties from distributions rather than calculating every photon scattering, thereby reducing computational demand.

\end{abstract}

\keywords{}

\section{Introduction}
\label{sec:intro}

Given the abundance of hydrogen in the universe, the Lyman $\alpha$ (Ly$\alpha$) line is an important component of radiation fields in a wide range of astrophysical settings. \lya radiation transport is an active area of research in the study of planets, stars, galaxies, and cosmology \citep{2019SAAS...46....1D}. An example application motivating our work is the role of \lya in planetary atmospheres. The outer layers of the atmosphere are central to a planet's evolution, since they can shelter the lower atmosphere from high energy radiation as well as regulate the escape of gas into space.  There are two sources of Ly$\alpha$: the star, and recombinations in the planet's atmosphere. \lya may ionize atoms and dissociate molecules, as well as exert pressure forces that drive an outflow \citep{2018A&A...620A.147B}. \lya can also excite H atoms to the 2p state, creating a population of Balmer-line absorbers that can be observed via transmission spectroscopy \citep{2017ApJ...851..150H, 2021ApJ...907L..47Y}. Due to the low gas densities in the upper atmosphere, collisional de-excitation and broadening are of secondary importance and \lya may undergo ``resonant scattering''.

Hubble Space Telescope (HST) observations with the STIS have found large \lya transit depths around a handful of exoplanets \citep{2003Natur.422..143V, 2012A&A...543L...4L, 2012A&A...547A..18E, 2015Natur.522..459E,  2017A&A...597A..26B, 2017A&A...599L...3B, 2017A&A...602A.106B, 2018A&A...620A.147B, 2019AJ....158...50W, 2019EPSC...13.1928L, 2020ApJ...888L..21G,2021arXiv210309864B}. These observations have revealed a population of atoms extending out to distances of order a few planetary radii or more for several planets around bright, nearby stars, motivating a study of the physics of \lya interactions with the H atom population. The transition from the atomic to the molecular layer in these hot upper atmospheres may take place at pressures of order ${\sim}\ 10\ \mu$bar (see discussion in \citealt{2017ApJ...851..150H} for details). This suggests the presence of a thick layer of atomic H which can have a line center optical depth of $\tau_0\ {\sim}\ 10^8$. A careful treatment of resonant scattering is necessary in order to construct accurate models of H atom excitation, heating, and radiative forces. 

Due to the technical challenge of including resonant scattering, the fully three-dimensional geometry, and the presence of an outflow, numerical simulations may be required to fully understand the dynamics of these irradiated exoplanet atmospheres. The large optical depths at \lya line center impose a steep computational cost for solving radiative transfer with Monte Carlo methods directly coupled with fluid dynamics \citep{2017MNRAS.464.2963S}. The number of scatterings a photon undergoes is proportional to the line center optical depth, $\tau_0$, of the domain \citep{1972ApJ...174..439A}. Near the base of the atomic layer, the line center optical depth is ${\sim}10^8$, so most of the time is spent following photons in these cells. A method that can accurately characterize transfer through these zones without following every photon scattering has the potential to greatly accelerate the calculation (see e.g. \citealt{1968ApJ...153..783A,2002ApJ...567..922A}).

Approximate analytic solutions for resonant scattering exist in certain limits. \citet{1973MNRAS.162...43H} showed that when most of the radiation is in the damping wings, the transfer equation reduces to the Poisson equation. However, their solution uses an ansatz to handle the boundary condition. To our knowledge, the errors introduced by this treatment have never been quantified. They attempt a separation of variables as $J(\tau,\sigma) = \theta(\tau) j(\sigma)$ in spatial variable $\tau$ and frequency variable $\sigma$ (their Equations 16 and 23). The solutions for the eigenfunctions $\theta(\tau)$ and $j(\sigma)$ then depend explicitly on the separation constant $\lambda$. In order to satisfy the boundary conditions, the separation constant is shown to satisfy an eigenvalue equation of the form
\be
\lambda \tan(\lambda B) & = & \frac{3}{2} \phi \Delta,
\label{eq:evalue}
\ee
where $2B$ is the slab optical depth at line center, $\phi$ is the line profile, and $\Delta$ is the Doppler width. The key point is that the line profile depends on one of the coordinates: frequency. This causes the eigenvalues of the separation constant to be frequency-dependent. Thus, the separation ``constant'' is not constant, and the function does not satisfy the Poisson equation since the frequency derivatives will act on the separation ``constant", giving extra terms. In the limit of large optical depth $B$, they approximate the eigenvalues as $\lambda_n B \simeq \pi (n-1/2)$, which gives zero mean intensity at the surface. Their Equation 34 subsequently allows $\lambda$ to have a small deviation from the above expression, which is explicitly frequency dependent. This allows a nonzero intensity at the surface, but at the cost of rendering the separation of variables assumption invalid. Our treatment, using the correct boundary condition, quantifies the errors in this ansatz.

Several other works have followed \citet{1973MNRAS.162...43H}. \citet{1990ApJ...350..216N} extends the solution to media of intermediate optical depth, including the effects of scattering in the Doppler core of the line. \citet{2006ApJ...649...14D} generalize the same problem to spherical geometry, as is used here. \cite{2020MNRAS.497.3925L} generalize both the slab and sphere solutions to arbitrary power-law density and emissivity profiles. Each of these works, and several others \citep{2020ApJS..250....9S, 2021MNRAS.504...89T}, use either the same surface boundary condition and ansatz as \citet{1973MNRAS.162...43H}, or use a solution that does not handle the frequency-dependence of the boundary condition. Our novel steady-state solution involves a frequency-dependent correction to the solution that fixes an observed excess at the spectral peaks as compared with Monte Carlo, which is present in many of the works cited above.

The motivation for including time-dependence in the transfer equation is to characterize the distribution of photon escape times, which is needed to calculate the radiation moments in the Monte Carlo simulation. Additionally, steady-state solutions to this problem are not always sufficient to describe all the physics of \lya transport. Time-variable, optically-thick environments necessitate a time-dependent solution to include the dynamic effects of \lya transfer. These include the optical afterglow of gamma-ray bursts \citep{2010ApJ...716..604R} and \lya sources redshifted by cosmological expansion \citep{2011MNRAS.418..853X}, among others.

\section{STEADY-STATE SOLUTION}
\label{sec:steadystate}

Consider a sphere of radius $R$ with uniform density $n_{\rm sc}$, luminosity $L$, and line-center optical depth $\tau_0$, containing a point source of photons. We aim to find the intensity within the sphere as a function of radius and photon frequency. The point source is assumed to be a delta function in space and photon frequency. Photons of frequency $\nu$ near the line center frequency $\nu_0$ are considered. The photon frequency of the source is $\nu_s$. The Doppler width is $\Delta = \nu_0 v_{\rm th}/c$, where $v_{\rm th}=\sqrt{2k_{\rm B}T/m_{\rm H}}$ is the thermal speed of hydrogen atoms of mass $m_{\rm H}$ and temperature $T$, and $c$ is the speed of light. The photon frequency in Doppler units is $x = (\nu-\nu_0)/\Delta$, and $x_s = (\nu_s - \nu_0)/\Delta$ is the corresponding source frequency. For upper-state de-excitation rate $\Gamma$, the ratio of natural to Doppler broadening is $a=\Gamma/(4\pi \Delta)$. For the \lya transition and T=$10^4$ K, $a = 4.72\times 10^{-4}$. $\mathcal{H}(x,a)$ is the Voigt function, and the Voigt line profile is $\phi = \mathcal{H}(x,a)/(\sqrt{\pi} \Delta)$, which is normalized as $\int d\nu\, \phi(\nu) = 1$. The line center optical depth is $\tau_0 = kR/(\sqrt{\pi}\Delta)$, where $k = n_{\rm sc} \pi e^2 f/(m_e c)$. Here, $e$ and $m_e$ are the charge and mass of the electron, and $f$ is the oscillator strength of the transition, which is 0.4162 for \lya \citep{1986rpa..book.....R}.

Appendix \ref{app:rteqn_derivation} contains a derivation of the transfer equation for convenience. Starting with the full transfer equation, Equation (\ref{eq:finaleqn}), ignoring photon destruction and including a photon emission term given by Equation (\ref{eq:jem}), the steady-state transfer equation is
\be
\nabla^2 J + \left( \frac{k}{\Delta} \right)^2 \frac{\partial^2 J}{\partial \sigma^2} & = & 
- \frac{ \sqrt{6} kL}{4\pi \Delta^2} \delta^3(\vec{x} - \vec{x}_s) \delta (\sigma - \sigma_{\rm s}).
\label{eq:rt_no_destr}
\ee
where $J$ is the mean intensity, the spatial variable is $\vec{x}$, and $\vec{x}_s$ is the position of the source. We will consider only the case where $\vec{x}_s=0$. Following \citet{1973MNRAS.162...43H}, we have used a change of variables in photon frequency from $x$ to $\sigma$,
\be \label{eq:int_change_of_variables}
\sigma(x) = \sqrt{\frac{2}{3}}\int_0^x \frac{dx}{\phi(x) \Delta} \approx \sqrt{\frac{2}{3}}\frac{\pi}{a}\frac{x^3}{3}, 
\ee
where the approximation is applicable in the damping wing. From Equation (\ref{eq:app:line_profile_wing}), the line profile is then approximately 
\be \label{eq:line_profile_approx}
\phi \approx \frac{a}{\pi x^2 \Delta} \approx \frac{1}{3 \Delta}\left(\frac{2a}{\pi}\right)^{1/3}|\sigma|^{-2/3}.
\ee
In Equation (\ref{eq:rt_no_destr}), $\sigma_s \equiv \sigma(x_s)$ is the photon frequency of the source. $\sigma_s$ is interchangeable with $x_s$ and $\nu_s$ in Doppler widths or Hz, respectively. Balancing the two terms on the left-hand side of Equation (\ref{eq:rt_no_destr}) gives $\sigma \ {\sim}\ \tau_0$, or $x_{\rm peak}\ {\sim}\ (a\tau_0)^{1/3}$. The boundary condition of no incoming intensity at the surface \citep{1986rpa..book.....R} is
\be
J & = & \sqrt{3} H
\label{eq:bc}
\ee
at $r=R$. 

\begin{figure}
    \centering
    \includegraphics{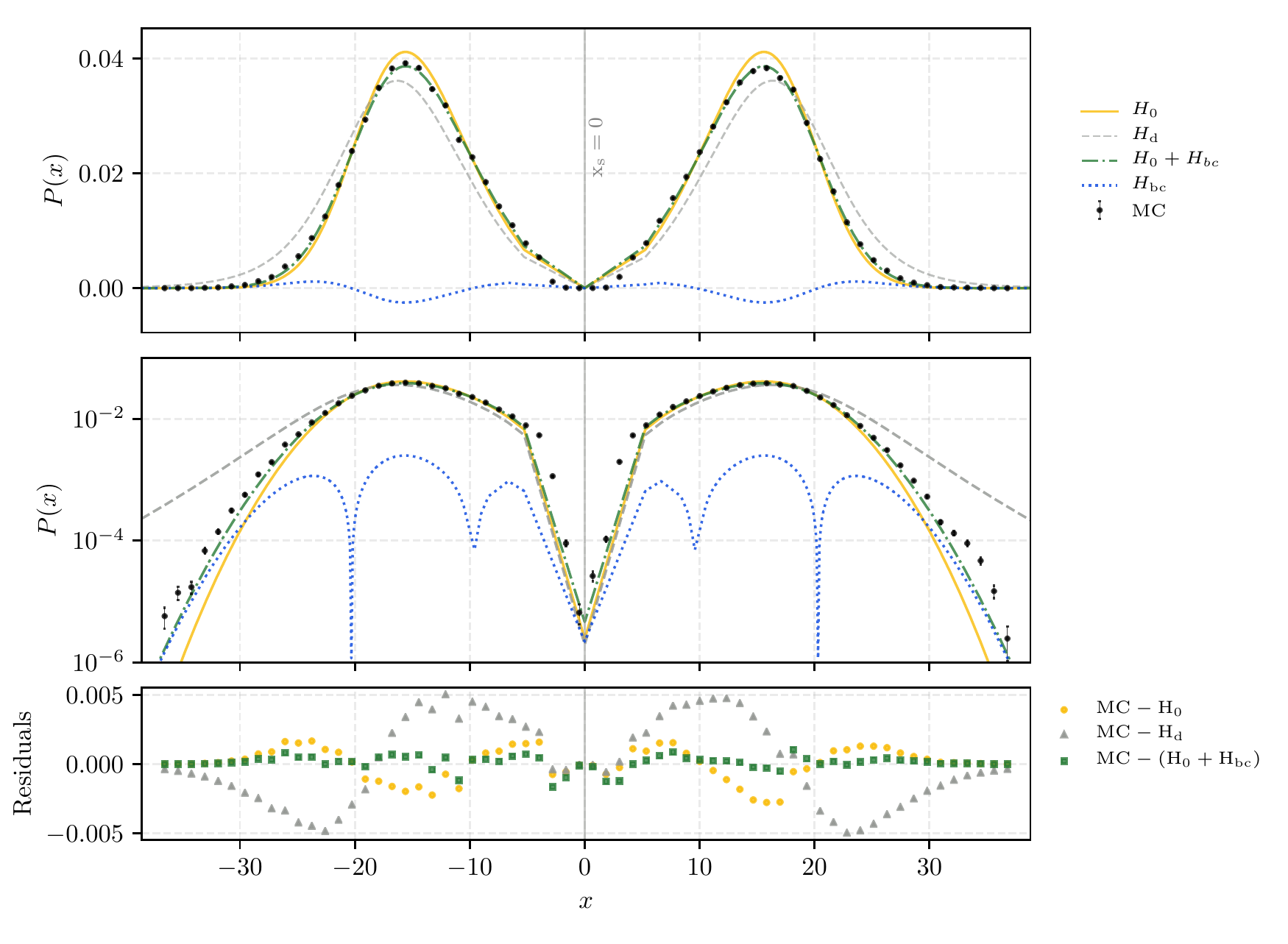}
    \caption{Spectrum $P(x)$ vs. frequency $x$ for $\tau_0 = 10^7$ with $x_s = 0$. At $T=10^4$ K, this corresponds to $a\tau_0=4.7 \times 10^3$. The legend to the right describes each line style. For reference, $\rm H_{0}$: fiducial solution, $\rm H_{d}$: divergent solution, $\rm H_{bc}$: our boundary condition correction to the fiducial solution. The top panel is linear scale, the middle panel is log scale with $|\rm H_{bc}|$ shown instead of $\rm H_{bc}$, and the bottom panel is the residual of each solution with Monte Carlo.} 
    \label{fig:sol_mc_residual_0}
\end{figure}

A solution for the mean intensity $J_d$ which is divergent at the origin
and $\sigma=\sigma_s$ and is zero at infinity is presented in \citet{1990ApJ...350..216N}. Here it is extended to spherical geometry and generalized to allow emission frequencies away from line center: 
\be
J_{\rm d} & = & 
\left(\frac{\sqrt{6}k^2L}{16\pi^3 \Delta^3}\right)\left(\frac{1}{(kr/\Delta)^2 + (\sigma - \sigma_{\rm s})^2}\right)
\label{eq:Jd}
\ee

\be
H_{\rm d} & = & - \frac{1}{3k\phi} \frac{\partial J_d}{\partial r}
=  \left( \frac{1}{3k\phi} \right) 
\left( \frac{ \sqrt{6}k^3L }{ 8\pi^3 \Delta^4} \right)
\left( \frac{k r/\Delta}{ \left[ (kr/\Delta)^2 + (\sigma-\sigma_{\rm s})^2 \right]^2 } \right).
\label{eq:Hd}
\ee
This solution is useful as a simple analytic formula. However, it is not a good approximation to the true solution, as it is too large at $r=R$ by a factor of $J_{\rm d}(R,\sigma)/ H_{\rm d}(R,\sigma) \sim a\tau_0/x^2 \sim (a\tau_0)^{1/3} \gg 1$ and does not adhere to the correct boundary condition. This solution is included in Figure \ref{fig:sol_mc_residual_0} for illustration.

A better approximation to the true solution has been derived by \citet{2006ApJ...649...14D}, who generalized the closed-form solution in slab geometry found in \citet{1973MNRAS.162...43H}. It satisfies a $J=0$ boundary condition at $r=R$. Again, we generalize their solution to allow emission at frequency $\sigma_{\rm s}$ away from line center. The result can be written as a sum over spatial modes,
\be \label{eq:J0_sum}
J_0 = \frac{\sqrt{6}L}{16\pi \Delta} \frac{1}{R^2}\sum_{n=1}^{\infty}n\frac{\sin{\kappa_n r}}{\kappa_n r}\exp{\left(\frac{-\kappa_n \Delta}{k}|\sigma - \sigma_s|\right)},
\ee
and
\be \label{eq:H0_sum}
H_0 = - \frac{1}{3k\phi} \frac{\partial J_0}{\partial r} = -\frac{1}{3k\phi}\frac{\sqrt{6}L}{16\pi\Delta} \frac{1}{R^2}\sum_{n=1}^{\infty}n\left(\frac{\cos{\kappa_n r}}{r} - \frac{\sin{\kappa_n r}}{\kappa_n r^2}\right)\exp{\left(\frac{-\kappa_n \Delta}{k}|\sigma - \sigma_s|\right)},
\ee
where $\kappa_n=n\pi/R$. These can be summed to give the closed form expressions
\be
J_0 & = & \frac{\sqrt{6}L}{32\pi^2 \Delta}
\frac{1}{Rr}
\left( 
\frac{ \sin(\pi r/R) }{ \cosh \left[ \frac{\pi \Delta}{k R} (\sigma - \sigma_s) \right] - \cos(\pi r/R)}
\right)
\label{eq:J0}
\ee
and
\be
H_0 & = &\frac{1}{3k\phi}
\frac{\sqrt{6}L}{32\pi^2 \Delta}
\frac{1}{Rr^2}
\left( 
\frac{ \sin(\pi r/R) }{ \cosh \left[ \frac{\pi \Delta}{k R} (\sigma - \sigma_s) \right] - \cos(\pi r/R)}
\right. \nonumber \\ & & \left. - \left( \frac{\pi r}{R} \right)
\frac{ \cos(\pi r/R) }{ \cosh \left[ \frac{\pi \Delta}{k R} (\sigma - \sigma_s) \right] - \cos(\pi r/R)}
+ \left( \frac{\pi r}{R} \right)
\frac{ \sin^2(\pi r/R) }{ \left[ \cosh \left[ \frac{\pi \Delta}{k R} (\sigma - \sigma_s) \right] - \cos(\pi r/R) \right]^2 }
\right).
\label{eq:H0}
\ee
These solutions agree with Equations (\ref{eq:Jd}) and (\ref{eq:Hd}) when the arguments of the trigonometric and hyperbolic functions are small. Again $J_0 \gg H_0$, except near $r=R$, where it goes to zero. The flux at $r=R$ can be written
\be
\nonumber
H_0(R, \sigma) & = & - \frac{1}{3k\phi}
\frac{\sqrt{6}L}{16\pi \Delta}
\frac{1}{R^3}
\sum_{n=1}^{\infty} 
n (-1)^n \exp{\left(\frac{-\kappa_n \Delta}{k}|\sigma - \sigma_s|\right)}\\
& = &  \frac{1}{3k\phi}
\frac{\sqrt{6}L}{32\pi \Delta}
\frac{1}{R^3}
\left( 
\frac{ 1 }{ \cosh \left[ \frac{\pi \Delta}{k R} (\sigma - \sigma_s) \right] +1 }
\right).
\label{eq:H0surf}
\ee
Equation (\ref{eq:H0}) will be shown to be a better approximation to the solution than Equation (\ref{eq:Hd}). It is still valid near the delta function at $r=0$, but is also a better approximation at $r=R$. $J_0$ decreases exponentially, rather than as a power-law in frequency as it does for $J_d$, giving a much smaller flux in the line wings as compared to the divergent solution. 

In order to enforce the boundary conditions, a different solution method is attempted here, namely a continuous Fourier expansion in the frequency variable $\sigma$. The solution of this problem is split into two pieces: $J_0$ which includes the delta function source and satisfies $J=0$ at $r=R$, and $J_{\rm bc}$ which allows the boundary condition $J=\sqrt{3}H$ to be satisfied at $r=R$. The total solution is
\be
J(r,\sigma) & = & J_0(r,\sigma) + J_{\rm bc}(r,\sigma)
\ee
and
\be \label{eq:totalflux}
H(r,\sigma) & = & H_0(r,\sigma) + H_{\rm bc}(r,\sigma).
\ee
The additional term $J_{\rm bc}$ must then be a solution of the homogeneous equation
\be \label{eq:diffeq}
\frac{\partial^2J_{\rm bc}}{\partial r^2} + \frac{2}{r} \frac{\partial J_{\rm bc}}{\partial r}
+ \left( \frac{k}{\Delta} \right)^2 \frac{\partial^2 J_{\rm bc}}{\partial \sigma^2} &= & 0
\ee
with no delta function source term, and it must allow the boundary conditions to be satisfied at the surface. Since $J_0(R,
\sigma)=0$, the surface boundary condition becomes
\be
J_{\rm bc}(R,\sigma) - \sqrt{3} H_{\rm bc}(R,\sigma) & = 
\sqrt{3} H_0(R,\sigma).
\label{eq:bc2}
\ee
Inserting a frequency dependence $J_{\rm bc} \propto e^{is\sigma}$, for ``wavenumber" $s$, gives the equation for modified spherical Bessel functions of the first kind, $i_0(z)=\sinh(z)/z$ for the radial dependence. The solution can then be represented as
\be
J_{\rm bc}(r,\sigma) & = & 
\int_{-\infty}^\infty \frac{ds}{2\pi} e^{is\sigma} A(s) 
\frac{i_0(krs/\Delta)}{i_0(kRs/\Delta)},
\label{eq:Jbc}
\ee
where $A(s)$ is the Fourier amplitude. Inserting Equation (\ref{eq:Jbc}) into Equation (\ref{eq:bc2}) leads to the following equation for the Fourier amplitudes,
\be
\int_{-\infty}^\infty \frac{ds}{2\pi} e^{is\sigma} A(s)
\left[ 1 + \left( \frac{s}{\sqrt{3} \Delta \phi} \right) \left( \frac{i_0^\prime(kRs/\Delta)}{i_0(kRs/\Delta)} \right) \right]
& = & \sqrt{3} H_0(R,\sigma).
\label{eq:bc3}
\ee
Discretization of Equation (\ref{eq:bc3}) for frequency variables $\sigma_i$ and wavenumbers $s_j$
leads to a set of coupled linear equations for the $A(s_j)$. We use equally-spaced points $\delta \sigma = 2\sigma_{\rm max}/(N-1)$ and $\delta s = 2\pi/(N\delta \sigma)$, where $N$ is the number of points for each grid. The maximum frequency is set as $\sigma_{\rm max} = {\rm constant} \times \tau_0$, for a large enough constant that the end of the frequency grid is at such small intensities that it does not affect the solution except close to the boundaries. The number of points was increased until the solution was well-resolved near line center, and only became inaccurate close to the boundaries. We found that values of $N=4097$ and $\sigma_{\rm max} = 60 \tau_0$ were sufficient. Given the Fourier amplitudes $A(s)$, $J_{\rm bc}$ is computed using Equation (\ref{eq:Jbc}), and the flux is given by
\be
H_{\rm bc}(r,\sigma) & = & -\frac{1}{3k\phi}
\frac{\partial J_{\rm bc}(r,\sigma)}{\partial r}
= -\frac{1}{3k\phi}
\int_{-\infty}^\infty \frac{ds}{2\pi} e^{is\sigma} A(s) 
\left( \frac{ks}{\Delta} \right) 
\left( \frac{i_0^\prime(krs/\Delta)}{i_0(kRs/\Delta)} \right).
\label{eq:Hbc}
\ee
The Bessel functions are finite at the center and rise steeply toward the surface when $kRs/\Delta \gg 1$. 

\subsection{Scaling with Line Center Optical Depth $\tau_0$}

We now estimate the scaling of $H_{\rm bc}$ with $\tau_0$. In the limit $J_{\rm bc} \gg H_{\rm bc}$, we find that $J_{\rm bc} \approx \sqrt{3} H_0$ from Equation (\ref{eq:bc2}). We  estimate $H_{\rm bc}$ from $J_{\rm bc}$ using Equation (\ref{eq:Hbc}) as
\be
H_{\rm bc}(R, \sigma) \approx \frac{1}{\sqrt{3}k\phi}\frac{ks}{\Delta}H_0\ {\sim}\ H_0 s \frac{x^2}{a}\ {\sim}\ H_0 \frac{1}{\tau_0}\frac{(a\tau_0)^{2/3}}{a}\ {\sim}\ H_0 (a\tau_0)^{-1/3},
\ee
where we have used $s\ {\sim}\ 1/\sigma\ {\sim}\ 1/\tau_0$ so that
\be \label{eq:hbc_scaling}
\frac{H_{\rm bc}(R, \sigma)}{H_0(R, \sigma)} \propto (a\tau_0)^{-1/3}.
\ee
At large $\tau_0$, it is expected that the correction term will be small, but it will become increasingly important as $\tau_0$ decreases. Our solution of the transfer equation is only valid when the peaks of the spectral energy distribution lie well outside of the Doppler core, i.e., for large $\tau_0$. The value of $x$ at which the Doppler and Lorentzian components of the line profile are equal is $x_{\rm cw}=3.3$. Setting $x_{\rm cw} = x_{\rm peak}$ and solving for $\tau_0$ gives the value at which the peak of the spectrum falls at the Doppler core boundary, which is $\tau_{\rm cp} \approx 10^5$ (``core-peak'' optical depth). Hence $H_{\rm bc}/H_0 {\sim} (\tau_{\rm cp}/\tau_0)^{1/3}$ is large at $\tau_0 \leq \tau_{\rm cp}$ and decreases relatively slowly as $\tau_0$ increases. Additionally, the optical depth at $x_{\rm peak}$ is proportional to $(a\tau_0)^{1/3}$, so photons here become optically thin when $a\tau_0 {\sim} 1$.

\subsection{Comparison to Monte Carlo}
The Monte Carlo method is used to solve the transfer equation numerically in order to compare the analytic approximation to an ``exact'' solution. This method is valid at all $\tau_0$, being restricted only by the computational demand, which grows proportionally to the number of photons used and $\tau_0$. For each simulation, a total of ${\sim}10^6$ photon packets are initialized at a monochromatic source frequency $x_s$ and are allowed to propagate through the sphere until escaping, at which point their positions, outgoing angles, and escape frequencies are tabulated to obtain the spectrum at the surface of the spherical simulation domain. A constant temperature of $T=10^4\ \rm K$ is set for the gas. Frequency redistribution is calculated at each scattering. In the comparisons shown in this section, the raw photon data is binned in frequency to obtain spectra. Further details of the Monte Carlo implementation are discussed in \citet{2017ApJ...851..150H}.

We now compare each of the previously-discussed solutions for surface flux to the Monte Carlo results. The spectrum $P(x)$ is defined as the specific luminosity at the surface divided by the source luminosity, or
\be \label{eq:prob_spectrum}
P(x) = \frac{16\pi^2R^2H(R, x)\Delta}{L}.
\ee
This is normalized so that $\int P(x)dx = 1$. Since $H(R, x)$ is per $d\nu$, a factor of $\Delta$ gives the expression the correct units. 

In Figure \ref{fig:sol_mc_residual_0}, the Monte Carlo spectrum is shown along with that of the solutions $H_{\rm d}$, $H_0$, and $H_0 + H_{\rm bc}$ for an optical depth of $\tau_0 = 10^7$ and photons emitted at line center $\rm x_s = 0$.  Note that the errorbars shown on the Monte Carlo data points are proportional to $\sqrt{N}$, with $N$ being the photon count in each frequency bin, since the photons are all equally weighted. The $H_{\rm bc}$ term is negative at the peak of the spectrum and positive in the line wing such that, when added to $H_0$, it corrects for the apparent excess of flux in the peaks of the spectrum. The solution with the correct frequency-dependent boundary condition enforced, $H_0 + H_{\rm bc}$, has lower residuals to Monte Carlo results than the other solutions, especially in the line wing. The boundary term corrects the deficit of $H_0$ in the line wings, further improving agreement with the numerical result. The residuals to the $H_0$ solution are a close match to the $H_{\rm bc}$ term, since the Monte Carlo represents the ``true'' solution, $H$, and $H_{\rm bc} = H - H_0$. It is evident that the divergent solution $H_{\rm d}$ fails in the line wings. Also note that the ``V'' shape of the solution in the line core is due to the low number of points plotted, as the analytic solutions are not valid in this frequency regime since they utilize the damping wing approximation of the Voigt line profile.

 \begin{figure}
    \centering
    \includegraphics[width=\textwidth]{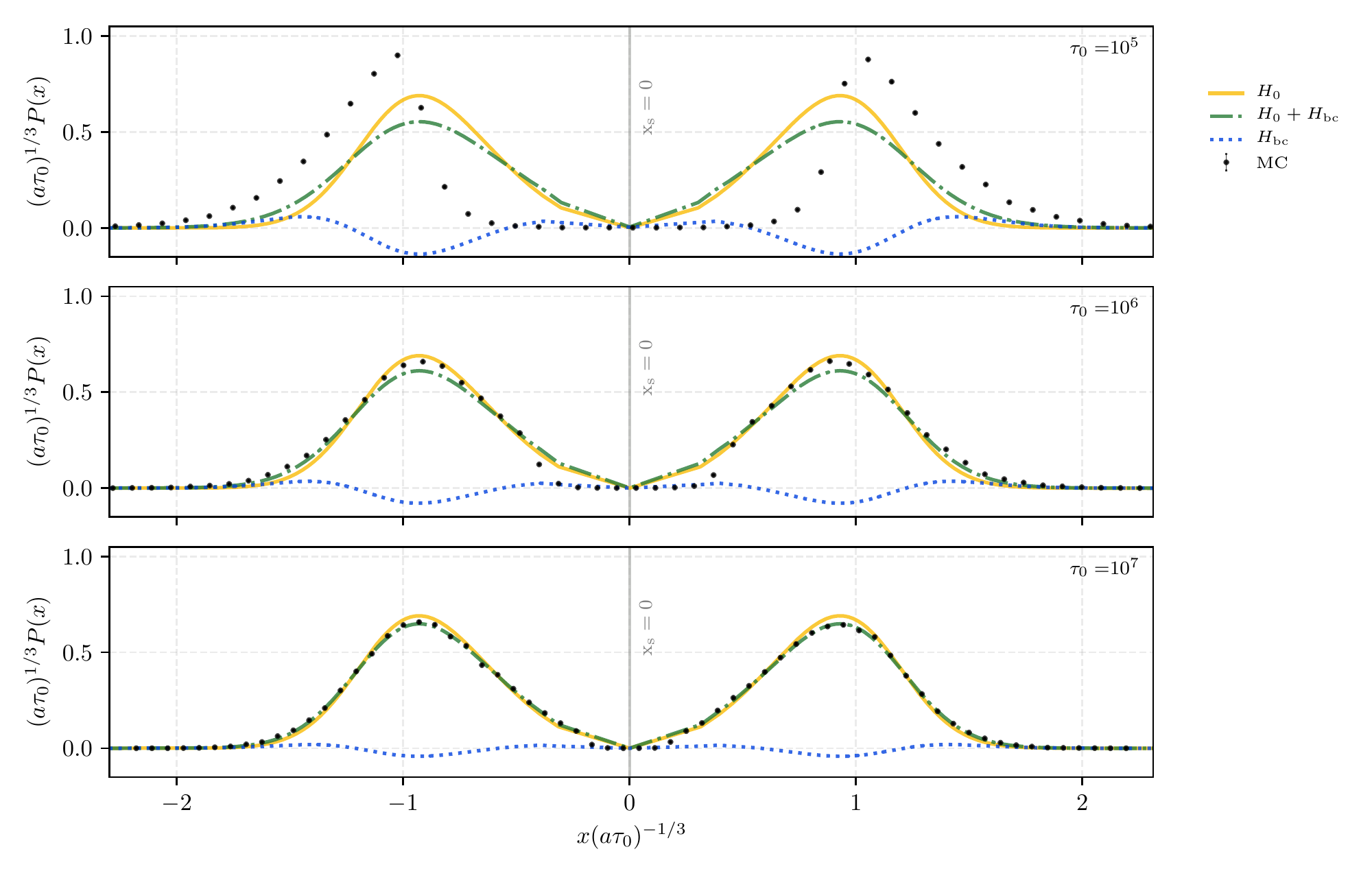}
    \caption{The same as Figure \ref{fig:sol_mc_residual_0}, but for $\tau_0 = 10^5$ (top panel), $10^6$ (middle panel), and $10^7$ (bottom panel). The $x$ and $y$ axes are scaled by $(a\tau_0)^{1/3}$.}
    \label{fig:sol_mc_tau}
\end{figure}

The size of $H_{\rm bc}$ is dependent on $\tau_0$. $H_{\rm bc}$ is significant even at $\tau_0 {\sim} 10^7$ where the $H_0$ solution is expected to perform well, i.e., photons are pushed further out into the wing where the simplifying assumptions made in the derivation of the differential equation are a better approximation. 

In Figure \ref{fig:sol_mc_tau} we show the solutions alongside Monte Carlo, now for three different optical depths $\tau_0=10^5, 10^6$, and $10^7$. From Equation (\ref{eq:hbc_scaling}), the size of the term $H_{\rm bc}$ should become smaller with larger optical depths, following a $(a\tau_0)^{-1/3}$ scaling. Indeed, agreement between Equation (\ref{eq:H0surf}) and the Monte Carlo points in Figure \ref{fig:sol_mc_tau} improves as $\tau_0$ increases, with $H_{\rm bc}$ providing a fractionally smaller correction to $H_0$. One factor of $(a\tau_0)^{1/3}$ has been scaled out of the x-axis such that the peaks of the distributions are horizontally aligned. This scaling has also been applied to the y-axis to preserve normalization of the escape probability. At lower $\tau_0$, the scattering of photons within the Doppler core of the line becomes important, but our analytic solution does not include this effect. The effects of line core scattering can be seen in the Monte Carlo data for $\tau_0=10^5$ and, to a lesser extent, $
\tau_0=10^6$.
 
 \begin{figure}
    \centering
    \includegraphics[width=\textwidth]{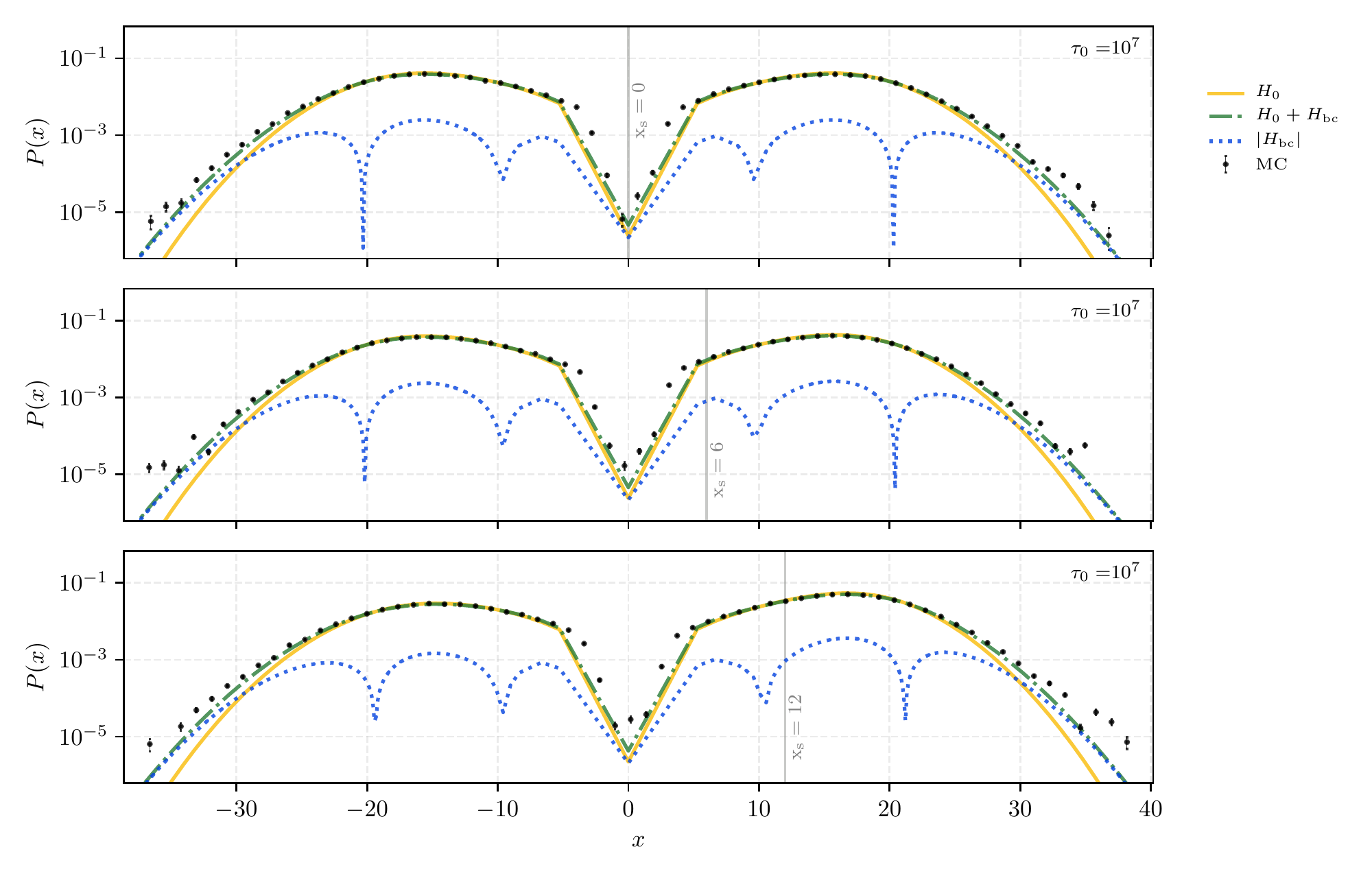}
    \caption{The same as Figure \ref{fig:sol_mc_residual_0}, but for $\tau_0=10^7$ and $\rm x_s = 0$ (top panel), $6$ (middle panel), and $12$ (bottom panel). The optical depth at each of these source frequencies is $\tau_s = 10^7, 77,$ and $19$, respectively.} 
    \label{fig:sol_mc_xinit}
\end{figure}

 \begin{figure}
    \centering
    \includegraphics[width=\textwidth]{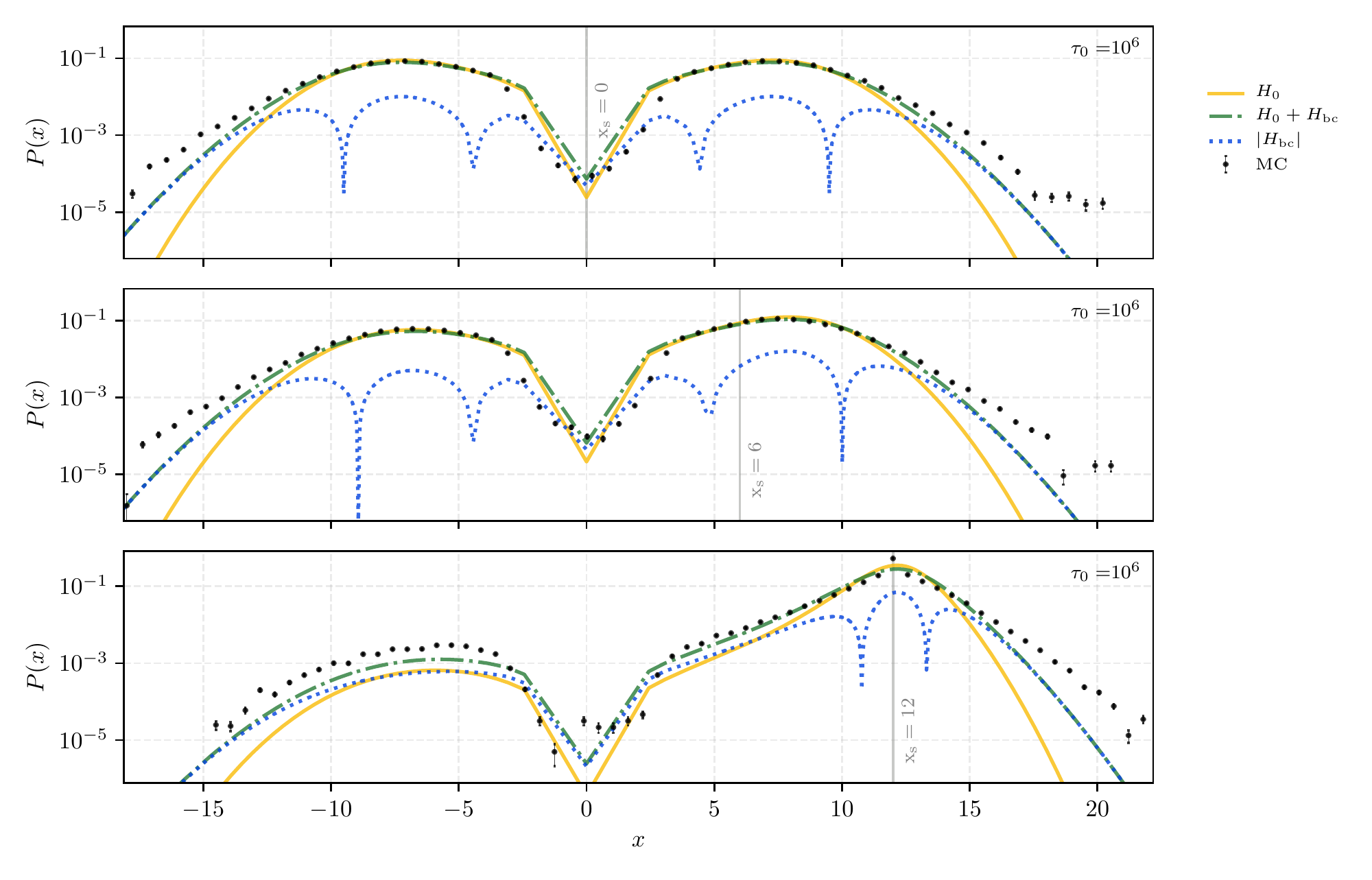}
    \caption{The same as Figure \ref{fig:sol_mc_xinit}, but at a lower optical depth $\tau_0 = 10^6$. The shift $x_s$ is a much larger fraction of the distance to the spectral peak $(a\tau_0)^{1/3}$, and thus the asymmetry in the spectrum is much larger. The optical depth at the source frequency is $\tau_s = 10^6$, $7.7$, and $1.9$ for $x_s=0, 6,$ and $12$, respectively. } 
    \label{fig:sol_mc_xinit_lowtau}
\end{figure}

Next, we show $P(x)$ for $x_s \neq 0$. Photons initialized further out in the line wing have larger mean free paths. The larger spatial diffusion implies greater escape probability for these photons. In the limit that $|\rm x_s|$ becomes large, the distribution becomes a delta function at $x_s$ as all photons escape the sphere without scattering. 

Figure \ref{fig:sol_mc_xinit} shows calculations performed for $x_s = 0, 6$, and $12$ and $\tau_0=10^7$. The asymmetry of the spectrum is slight for $\rm x_s = 6$ where $\tau(x_s)=77$, but is larger for $\rm x_s=12$ outside the line core where $\tau(x_s)=19$. It is seen here that the difference between the Monte Carlo data and $H_0$ becomes larger as $\rm x_s$ increases. Thus, for large $|x_s|$, inclusion of $H_{\rm bc}$ is more important.

Figure \ref{fig:sol_mc_xinit_lowtau} shows emission away from line center at the same values of $x_s$ as in Figure \ref{fig:sol_mc_xinit}, but for $\tau_0=10^6$ rather than $10^7$. The difference between the left and right side of the escaping spectrum is now substantial since $(a\tau_0)^{-1/3}$ increased by a factor of ${\sim}2$. It is clear from the figure that as $x_s$ extends further into the wing, the spectrum becomes more strongly peaked in frequency. Additionally, since the sphere is increasingly optically thin in the wing, we expect there to be a stronger disagreement with the Monte Carlo as the analytic solution assumed large optical depths.

\begin{figure}
    \centering
    \includegraphics[width=0.9\textwidth]{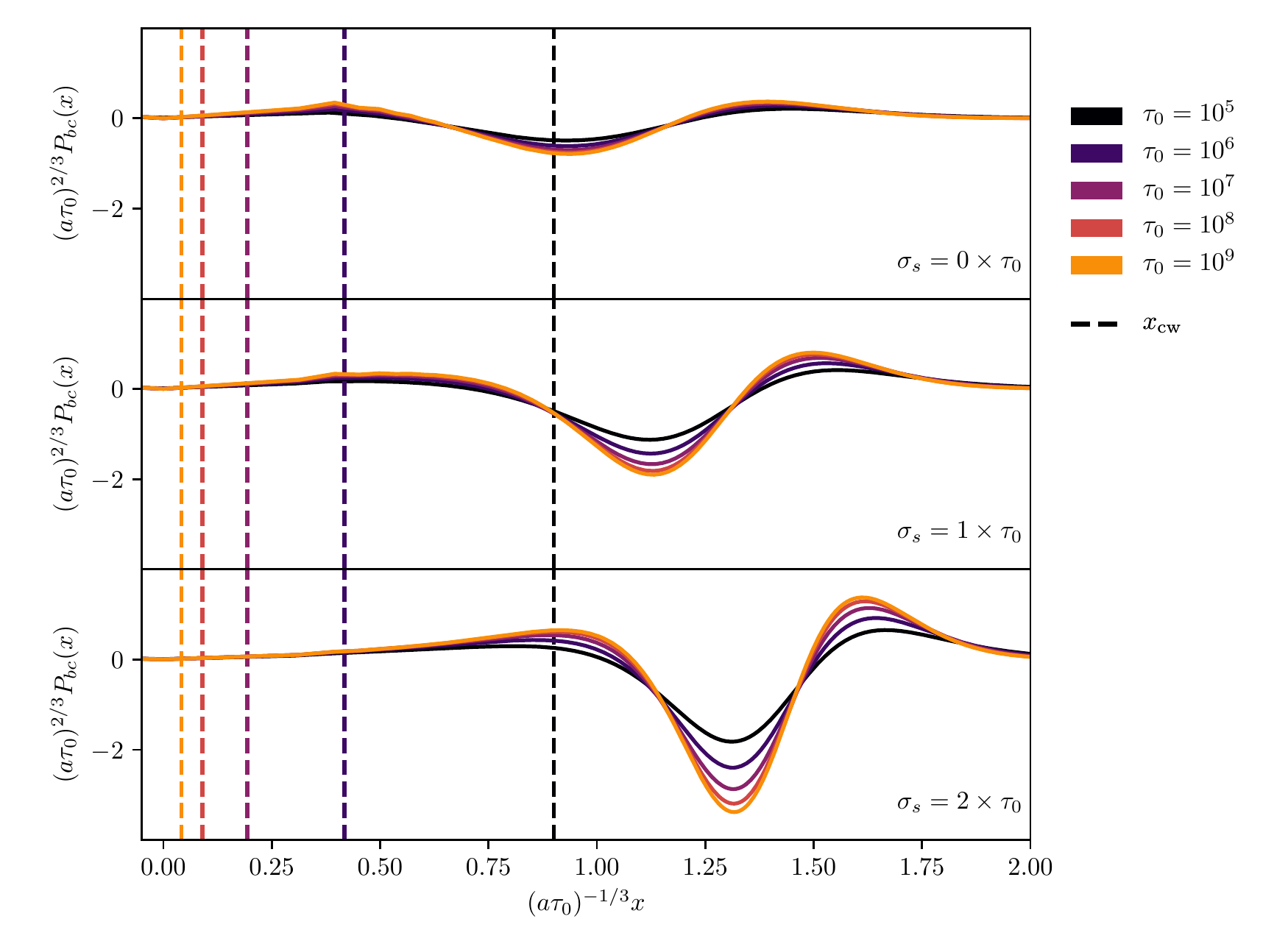}
    \caption{$P_{\rm bc}(x) = 16\pi^2R^2H_{\rm bc}(R, x)\Delta/L$ (Equation (\ref{eq:prob_spectrum}) with $H \rightarrow H_{\rm bc}$) vs. frequency $x$. Each panel shows $\sigma_s$ further from line center, labeled by $\sigma_s=0, \tau_0, 2\tau_0$. The solution is shown at a range of line center optical depths between $\tau_0=10^5$ and $10^9$ with solid lines. The location of the core-wing boundary $x_{cw}$ for each $\tau_0$ is shown with dashed vertical lines which match the color of the corresponding solution. The axes are scaled to show that the size of the correction factor agrees with the predicted $(a\tau_0)^{-1/3}$ scaling.}
    \label{fig:xinit}
\end{figure}

Figure \ref{fig:xinit} shows how the correction $P_{\rm bc}(x)$, Equation (\ref{eq:prob_spectrum}) with $H \rightarrow H_{\rm bc}$, scales with both $\tau_0$ and $x_s$. In this figure, $\sigma_s$ is shifted by integer multiples of $\tau_0$ (Equation \ref{eq:change_of_variables}) in each panel such that the source falls near the peak of the spectrum for each $\tau_0$. For clarity, only the $x > 0$ side of the spectrum is shown. From Equation (\ref{eq:hbc_scaling}), it is expected that the fractional size of $H_{\rm bc}$ relative to $H_0$ should become smaller with larger optical depths following $(a\tau_0)^{-1/3}$. This factor has been scaled out of the figure such that solutions for different $\tau_0$ and the same $\sigma_s$ should show close agreement in scale on the figure's vertical axis if the relation holds. Indeed, the scaled solutions converge as $\tau_0$ becomes larger, indicating agreement with the $(a\tau_0)^{-1/3}$ scaling. The remaining discrepancy present in the vertical axis for fixed $\sigma_s$ results from $x_{\rm peak}$ becoming close to $x_{\rm cw}$; at lower $\tau_0$, this causes the line profile approximation in the wing, Equation (\ref{eq:app:line_profile_wing}), to break down. From this, we conclude that the errors introduced by the incorrect separation of variables in \citet{1973MNRAS.162...43H}, \citet{1990ApJ...350..216N}, \citet{2006ApJ...649...14D} and others are indeed proportional to $(a\tau_0)^{-1/3}$.

\section{Time-Dependent Diffusion}
\label{sec:time_dependent}

In order to understand how long it takes for the photons to escape the uniform sphere of gas we must reintroduce the time dependence of the diffusion equation, which was ignored in the steady-state calculations in Section \ref{sec:steadystate}. To obtain the radiative intensity $I=dE/(dAdtd\Omega d\nu)$ on timescales comparable to the light-crossing time $t_{\rm lc} = R/c$, the time-dependent response to a delta function impulse is found. This allows the distribution of photon escape times (the ``wait time distribution'') to be characterized. For simplicity, a $J=0$ boundary condition will be used in the following derivations, which is a rough approximation for $a\tau_0 \gg 1$. 

\subsection{Derivation of the time-dependent solution}
\label{subsec:time_dependent:background}

The emissivity for an impulsive source with energy E, source position $\vec{x}_s$, and frequency $\nu_s$ is derived in Appendix \ref{app:rteqn_derivation}. Considering a photon source at $\vec{x}_s=0$, we have (Equation \ref{eq:jem})
\be
j_{\rm em} & = & \frac{E}{4\pi} \delta^3(\vec{x}) \delta(\nu-\nu_{\rm s})\delta (t) ,
\label{eq:jem2}
\ee 
The resulting equation for $J(r,\sigma,t)$ is
\be
 \frac{-3k\phi}{c} \frac{\partial J}{\partial t} + \nabla^2 J + \left( \frac{k}{\Delta} \right)^2 \frac{\partial^2 J}{\partial \sigma^2}
& = & - \frac{\sqrt{6} kE}{4\pi \Delta^2} \delta^3(\vec{x}) \delta (\sigma - \sigma_s ) \delta (t).
\label{eq:diffusion_eqn}
\ee
We employ an expansion in terms of spherical Bessel functions in $r$ and Fourier transform in time. The zeroth spherical Bessel function is $j_0(x) = \sin{x}/x$. The expansion for $J(r, \sigma, t)$ is then
\be
\label{eq:jrsigmat_expansion}
J(r, \sigma, t) = \sum_{n=1}^{\infty} \int_{-\infty}^\infty \frac{d\omega}{2\pi} e^{-i\omega t} j_0\left(\kappa_n r\right) J(n, \sigma, \omega),
\ee
with
\be \label{eq:jnsigmaomega}
J(n, \sigma, \omega) = \frac{2\kappa_n^2}{R} \int_0^R dr\ r^2 j_0(\kappa_n r) \int_{-\infty}^\infty dt\ e^{i\omega t} J(r, \sigma, t).
\ee
Here, $\kappa_n = n\pi/R$ and $\omega$ describes the time-dependence of $J$. Though it is written as a function of the photon frequency variable $\sigma$, $J(r, \sigma, t)$ is the specific mean intensity $dE/(dA dt d\nu)$ and is a distribution in $\nu$. The Fourier coefficient $J(n, \sigma, \omega)$ has units $dE/(dA d\nu)$. Using Equation (\ref{eq:jnsigmaomega}), we obtain
\be \label{eq:diffusion_plugged_in}
 \left( \frac{3k\phi}{c}i\omega  -   \kappa_n^2 \right) J(n,\sigma,\omega)  &+& \left( \frac{k}{\Delta} \right)^2 \frac{\partial^2J(n,\sigma,\omega)}{\partial\sigma^2} = -\frac{2\kappa_n^2}{R} \frac{\sqrt{6}}{4\pi} \frac{kE}{\Delta^2} \frac{1}{4\pi} \delta(\sigma - \sigma_s).
\ee
At $\sigma=\sigma_s$, continuity must be enforced,
\be \label{eq:matching_condition_1}
J(n, \sigma^-, \omega) = J(n, \sigma^+, \omega),
\ee
and the discontinuity in $dJ/d\sigma$ due to the source is
\be \label{eq:matching_condition_2}
\frac{\partial J(n, \sigma^+, \omega)}{\partial \sigma} - \frac{\partial J(n, \sigma^-, \omega)}{\partial \sigma} & = & 
- \frac{\sqrt{6}}{8} n^2 \frac{E}{kR^3}.
\ee
At large values of $\sigma$ the line profile $\phi$ is small and Equation (\ref{eq:diffusion_plugged_in}) becomes
\be \label{eq:diffusion_at_large_sigma}
\frac{\partial^2J}{\partial\sigma^2} \approx \frac{\Delta^2\kappa_n^2}{k^2} J,
\ee
which has solutions 
\be
J(n, \sigma, \omega)\ {\sim}\ e^{\pm \kappa_n \sigma \Delta / k}.
\ee
This approximate solution implies a boundary condition at large $|\sigma|$
\be \label{eq:single_j_derivative}
\frac{\partial J}{\partial \sigma} = \mp \frac{\kappa_n\Delta}{k} J,
\ee
where a negative sign is taken for large $+\sigma$ and a positive sign is taken for large $-\sigma$ to choose the finite solution as $|\sigma|\to \infty$. Numerical integrations are performed inward toward $\sigma_s$ over several domains: from large $|\sigma|$ to $\sigma_s$, from large $|\sigma|$ to 0, and from 0 to $\sigma_s$, depending on whether $\sigma_s$ is positive or negative. If $\sigma_s=0$, just two integrations are performed inward from large $|\sigma|$ to 0. Initial values for integration are obtained either by setting $J=1$ and $dJ/d\sigma$ from Equation (\ref{eq:single_j_derivative}) at large $|\sigma|$ or by matching $J$ and $dJ/d\sigma$ at 0. This gives $J$ and $J'$ on either side of $\sigma_s$, where a prime indicates the derivative $\partial/\partial \sigma$. By enforcing the matching conditions, Equations (\ref{eq:matching_condition_1}) and (\ref{eq:matching_condition_2}), the eigenfunctions $J(n, \sigma, \omega)$ are obtained over the domain of photon frequencies $\sigma$. Since the solutions are linear in the starting conditions, only two integrations with different starting values are necessary.

We now wish to reconstruct the specific mean intensity $J(r, \sigma, t)$. While one might expect this could be expressed as a sum over eigenmodes, the analysis presented in Appendix \ref{app:wkb} suggests this treatment is incomplete in the case where $x_s \neq 0$ and the solution is asymmetric about the line center. This ansatz does, however, roughly agree with Monte Carlo results for $x_s=0$ based on numerical calculations of this result.

Let us define the damping rate to be $\gamma \equiv i\omega$, which is real and positive for damped solutions. At the eigenvalues $\gamma=\gamma_{nm}$, the response $J(n, \sigma, \omega)$ is resonant. We find that near these $\gamma_{nm}$ poles an approximate expression for the resonant response of the eigenfunctions is
\be \label{eq:jnsigmaomega_approx}
J(n,\sigma,-i\gamma) & \simeq \frac{ J_{nm}(\sigma) }{\gamma_{nm} - \gamma} + C(\gamma, \sigma),
\ee
where $C(\gamma, \sigma)$ varies slowly in $\gamma$. If the $\omega$-integral in Equation (\ref{eq:jrsigmat_expansion}) could be closed at infinity and evaluated using the residue theorem, the result would be
\be
J(r,\sigma,t) & \simeq & j_0(\kappa_n r) J_{nm}(\sigma) e^{-\gamma_{nm}t}.
\ee
Summing over all spatial modes $n$ and over all eigenmodes $m$ for a given $n$, we obtain
\be \label{eq:Jrsigmat}
J(r,\sigma,t) & = & \sum_{n=1}^\infty j_0(\kappa_n r)  \sum_{m=1}^{\infty} J_{nm}(\sigma) e^{-\gamma_{nm}t}.
\ee
This ansatz captures the contributions from $n \times m$ simple poles. Taking a derivative with respect to $r$ and evaluating at the surface $r=R$, we use
\be
\frac{dj_0(\kappa_n r)}{dr} \bigg\rvert_R & =& \frac{d}{dr} \left[ \frac{\sin(\kappa_n r)}{\kappa_n r} \right]\bigg\rvert_R
=  \left( \frac{\cos(\kappa_n R)}{R} - \frac{\sin(\kappa_n R)}{\kappa_n R^2} \right)  =  \frac{(-1)^n}{R}
\ee
to obtain the flux, which is
\be
F(R,\sigma,t) & =& - \frac{4\pi}{3k\phi} \frac{dJ(R,\sigma,t)}{dr} 
= - \frac{4\pi}{3k\phi R}  \sum_{nm} (-1)^n J_{nm}(\sigma) e^{-\gamma_{nm}t}.
\ee
Multiplying by $4\pi R^2$ gives the energy per time per frequency emerging from the sphere to be
\be
\frac{dE}{dtd\nu} & = & - \frac{16\pi^2 R}{3k\phi}  \sum_{nm} (-1)^n J_{nm}(\sigma) e^{-\gamma_{nm}t}.
\label{eq:dEdtdnu}
\ee
Integrating over time yields a factor $1/\gamma_{nm}$, and by integrating over $d\nu$ we find
\be \label{eq:sum_rule}
E & = &  \sqrt{ \frac{3}{2} } \frac{16\pi^2R\Delta^2}{3k} \sum_{nm} (-1)^{n+1} \gamma_{nm}^{-1} \int d\sigma J_{nm}(\sigma).
\ee
This non-trivial ``sum rule'' provides a check on the values of $\gamma_{nm}$ and $J_{nm}(\sigma)$. This expression can also be written as
\be
1 & =& \sum_{nm} P_{nm},
\label{eq:sumrule}
\ee
where the contribution of each mode is
\be \label{eq:pnmsoln}
P_{nm} & \equiv & \sqrt{ \frac{3}{2} } \frac{16\pi^2R\Delta^2}{3kE}  (-1)^{n+1} \gamma_{nm}^{-1} \int d\sigma J_{nm}(\sigma).
\ee
These coefficients $P_{nm}$ are negative for odd values of $n$ and positive for even $n$. The size of each contribution scales roughly as $0.5/(m-7/8)^{2/3}$, with a weak dependence on $n$. This indicates the need for a large number of $n$ and $m$ to converge, in that it takes roughly ten times as many $m$ modes for a given $n$ to reduce the size of $P_{nm}$ by a factor of ${\sim}$5. The physical intuition for the convergence of these terms is that the $n$ spatial terms must provide sufficient spatial resolution to resolve the steep falloff in intensity at the surface of the sphere. Additionally, the function falls off steeply in frequency in the line wing, which requires more $m$ terms in the Fourier sum to resolve (also see the discussion of Figure \ref{fig:steadystate} in Section \ref{subsec:steadystatemc}). 

\subsection{Numerical calculation}

We seek now to calculate the eigenmodes $J_{nm}(\sigma)$ and eigenfrequencies $\gamma_{nm}$ for a given spatial $n$. These will be labelled by an index $m=1, 2, ...$. To measure the size of the response to detect where resonances occur, we sum the absolute value of $J(n,\sigma,-i\gamma)$ over the array $\sigma$. We call this response $f$, and use the index $j$ to represent the value of the response at discrete points $\gamma_j$ over a range of $\gamma$. In places where $f_j > f_{j-1}$ and $f_j>f_{j+1}$, we have bracketed a resonance that occurs in the interval $(\gamma_{j-1},\gamma_{j+1})$. To refine the value of the eigenfrequency before continuing the sweep in $\gamma$, we evaluate $f_{j-1}$, $f_j$, and $f_{j+1}$ at the points $(\gamma_{j-1},\gamma_{j},\gamma_{j+1})$. Assuming the form in Equation (\ref{eq:jnsigmaomega_approx}), a guess at the correct eigenvalue $\gamma_{nm}$ can be calculated by linear interpolation from
\be
\gamma_{\rm guess} &=& \frac{b\gamma_{j-1} - \gamma_{j+1}}{b - 1},
\ee
where
\be
b &=& \left(\frac{f_{j} - f_{j-1}}{f_{j} - f_{j+1}}\right)\left(\frac{\gamma_{j+1}-\gamma_{j}}{\gamma_{j-1}-\gamma_{j}}\right).
\ee
The error of the current guess is $|\gamma_{\rm guess} - \gamma_{j}|$. This error is reduced iteratively by replacing initial points $(\gamma_{j-1},\gamma_{j},\gamma_{j+1})$ with closer estimates while the size of the response grows as the resonance is approached. After iterating an eigenvalue $\gamma_{nm}$ to convergence, we now find the corresponding eigenfunction $J_{nm}(\sigma)$. We evaluate Equation \ref{eq:jnsigmaomega_approx} at two points $\gamma_1$ and $\gamma_2$ near the resonance, subtracting them and solving for $J_{nm}(\sigma)$ to find
\be
J_{nm}(\sigma) & \simeq & \frac{ J(n,\sigma,-i\gamma_1)  - J(n,\sigma,-i\gamma_2) }{ 1/(\gamma_{nm}-\gamma_1) - 1/(\gamma_{nm}-\gamma_2)},
\ee
\noindent where  $C(\gamma, \sigma)$ has cancelled in the difference.

\begin{figure}
    \centering
    \includegraphics{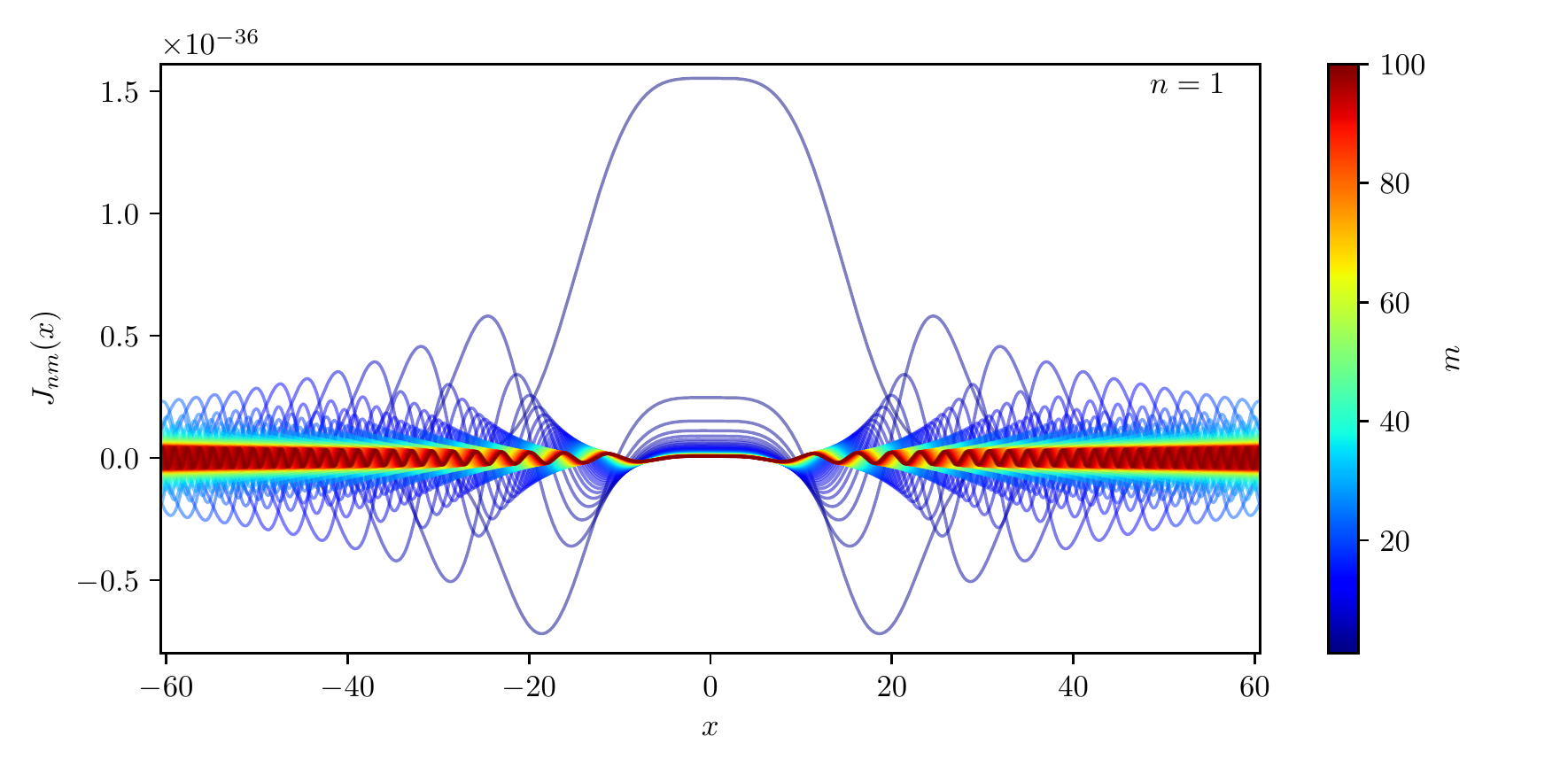}
    \caption{Eigenfunctions $J_{nm}(x)$ for the lowest-order spatial eigenmode $n=1$, and $m=1, ..., 100$ with $x_s=0$ and $\tau_0=10^7$. The scale of the $J_{nm}(\sigma)$ are set by the factor $E/(kR^3)$, which here is ${\sim}10^{-37}$ in ergs/(cm$^2$ Hz).}
    \label{fig:jsoln}
\end{figure}

The form of a single eigenmode $J_{nm}(\sigma)$ is oscillatory out to some turning point, $\sigma_{\rm tp}$, at which point the function becomes evanescent. The location of the turning point can be found by ignoring the delta-function discontinuity at the source frequency $\sigma_s$ in Equation (\ref{eq:diffusion_plugged_in}) and examining the resulting homogeneous differential equation. We obtain
\be \label{eq:wkb_differential_eqn}
\frac{d^2J}{d\sigma^2} & = & \left[ \left( \frac{\kappa_n \Delta }{k} \right)^2 - \frac{3\phi \gamma\Delta^2}{ck}\right] J,
\ee
where the line profile is approximated as in Equation (\ref{eq:app:line_profile_wing}). When the coefficient on the right hand side is positive, exponential growth or decaying evanescent solutions are found. This occurs in the line wings. When the coefficient on the right hand side is negative, oscillatory solutions are found (propagation), which occurs near the line core. The boundary between propagation and evanescence occurs at the turning point, given by
\be \label{eq:sigma_tp}
\sigma_{\rm tp} & = & \sqrt{\frac{2a}{\pi}}\left( \frac{k \gamma}{ \kappa_n^2 c \Delta} \right)^{3/2}
\ee
Thus, to ensure accuracy in each term of Equation (\ref{eq:Jrsigmat}), the bounds of $\sigma$ must be set sufficiently far outside of $\sigma_{\rm tp}$ such that the function is small at the edges. The scale of an $e$-folding in $J_{nm}(\sigma)$ is $k/(\kappa_n \Delta) = \tau_0 / (\sqrt{\pi} n)$, so a grid of $\sigma$ is chosen that spans a large enough number of $e$-foldings that no oscillatory behavior is present at the boundaries of the domain.

The eigenfunction's oscillatory forms have varying amplitudes which sum in Equation (\ref{eq:Jrsigmat}) to create the final form of the mean intensity. The largest contribution at late times always comes from the ($n=1, m=1$) lowest-order eigenfunction. Figure \ref{fig:jsoln} shows a set of eigenfunctions $J_{nm}(\sigma)$ to illustrate their relative scales for different $m$ at a fixed spatial eigenmode $n$. The overall scale of the $J_{nm}(\sigma)$ are set by the factor $E/(kR^3)$ with $E$ arbitrarily set to 1. For H atoms with $T=10^4$ K and $\tau_0=10^7$, an eigenfunction has typical size ${\sim} E a / \left(R^2 \Delta \right) = 10^{-37}$ in units of specific mean intensity times time. Additional terms add smaller-magnitude, faster-oscillating components that lead to higher accuracy upon summation with the lower-order terms in Equation (\ref{eq:Jrsigmat}). The oscillations of various modes must cancel in the Fourier sum, so many modes $m$ and $n$ are required for convergence to the solution. 
\begin{figure}
    \centering
    \includegraphics[width=1.0\textwidth]{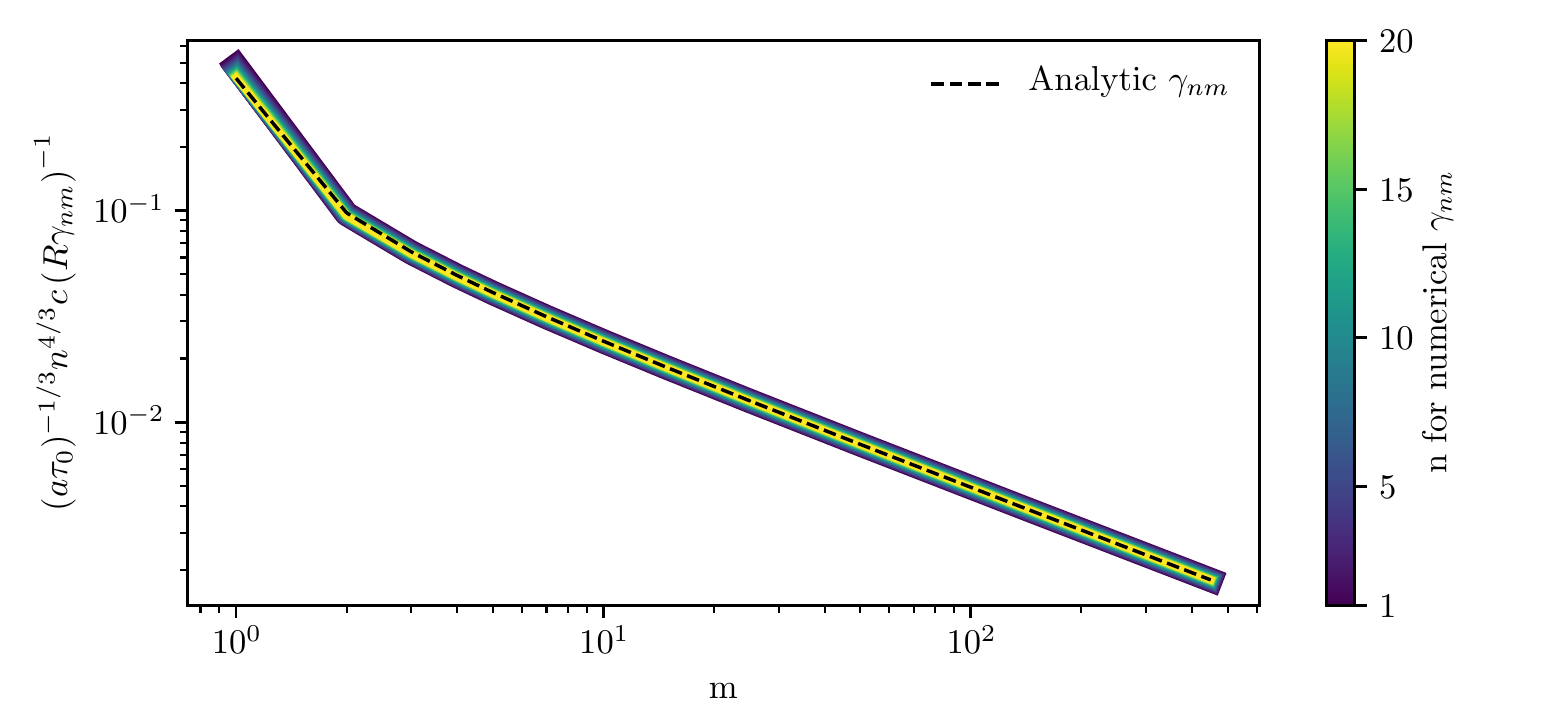}
    \caption{Dimensionless decay time vs. mode number. Overlapping solid lines are plotted for each $n$. The thickness of the lines decrease with $n$ in order to clearly show where they are in agreement. These numerically-obtained resonant frequencies are compared with the analytic expression in Equation (\ref{eq:gamma_nm}), shown as a dashed line. The parameters for the calculation are $\tau_0=10^7$ and $x_s=0$. The value of $n$ is given by the color bar on the right-hand side.}
    \label{fig:gamma_nm}
\end{figure}
The values of the $\gamma_{nm}$ can be described approximately with Equation (\ref{eq:gamma_nm}). Their values depend on $m$, $n$, and other physical parameters according to 
\be \label{eq:gamma_nm}
\gamma_{nm} = 2^{-1/3} \pi^{13/6} n^{4/3}\left(m-\frac{7}{8}\right)^{2/3}\frac{c}{R}(a\tau_0)^{-1/3}
\ee
as shown by setting the denominator of Equation \ref{eq:a3/a2} to zero in WKB approximation of Appendix \ref{app:wkb}. The power law in $m$ is weak, requiring up to $m=1000$ to reduce the scale of $\gamma_{nm}^{\ \ -1}$ by two orders of magnitude. When sweeping through to find resonances, Equation (\ref{eq:gamma_nm}) is used to set the scale of the sweep points $\gamma_j$ to ensure no $\gamma_{nm}$ are missed. The close agreement with the analytic expression shown in Figure \ref{fig:gamma_nm} indicate that the numerical solutions are accurate.

\subsection{Comparison with Steady State and Monte Carlo} \label{subsec:steadystatemc}

We now calculate the wait time distribution for escape from the sphere. This is obtained by integrating Equation (\ref{eq:dEdtdnu}) over all frequencies. We find
\be
P(t)  & = & \sqrt{\frac{3}{2}} \frac{16\pi^2 R \Delta^2 }{3kE}     \sum_{nm} (-1)^{n+1}  e^{-\gamma_{nm}t} \int d\sigma J_{nm}(\sigma) 
\nonumber \\ & = &  \sum_{nm} P_{nm} \gamma_{nm} e^{-\gamma_{nm}t},
\label{eq:waittime}
\ee
which is normalized to unity. For a sufficiently large number of spatial modes $n$ and frequency modes $m$, the result of this sum can agree with Monte Carlo escape time distributions when $x_s=0$. The late-time distribution is simply an exponential falloff. The rate constant of the falloff is the lowest-order eigenfrequency, $\gamma_{11}$, and its scale is determined by the coefficient $P_{11}$ as in Equation (\ref{eq:pnmsoln}). Thus, an approximate ``fitting function'' that captures both the peak of the escape time distribution and the exponential falloff is
\be \label{eq:fitting_function}
P(t) = \exp{\left[-\left(\frac{t_{\rm diff}}{t}\right)^2\right]} \times \gamma_{11} P_{11} e^{-\gamma_{11}t}.
\ee
The first term represents the early-time distribution, which then transitions to an exponential falloff past a point $ct_{\rm diff}/R = (a\tau_0)^{1/3}$, where $t_{\rm diff}$ is the characteristic diffusion timescale.

\begin{figure}
    \centering
    \includegraphics[width=0.75\textwidth]{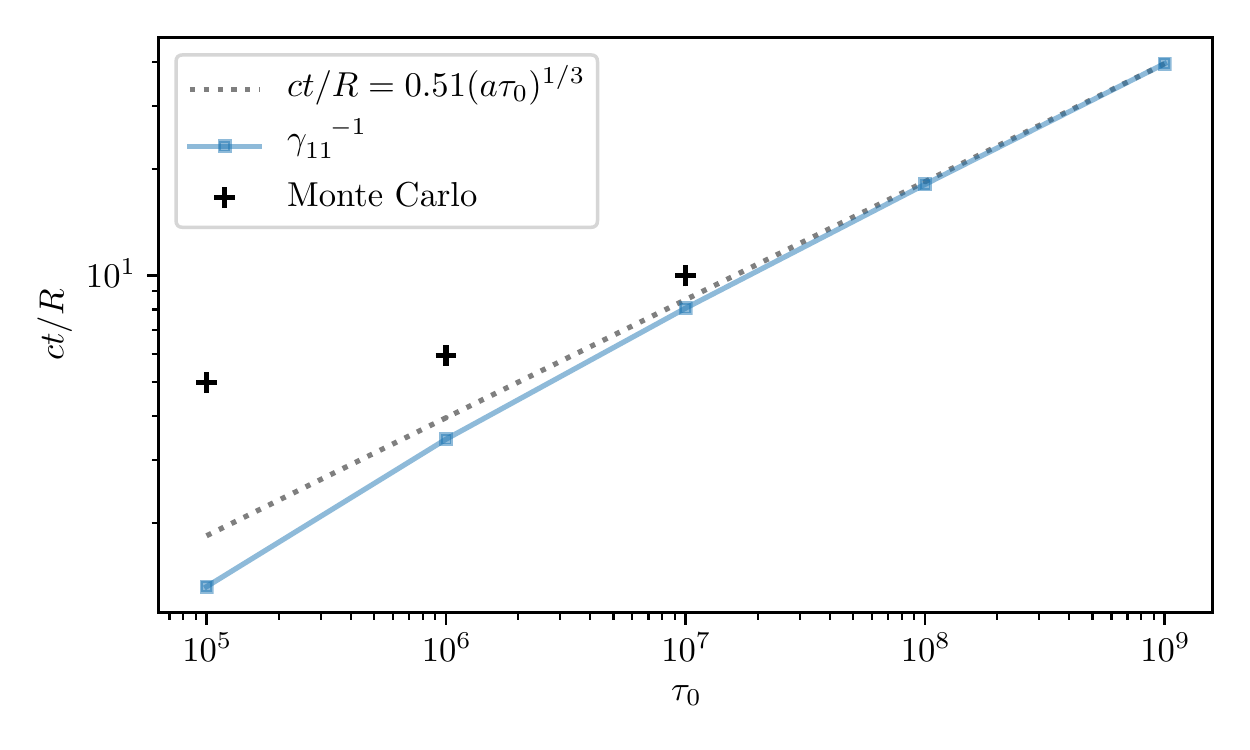}
    \caption{Late-time exponential decay timescale as a function of $\tau_0$. The Monte Carlo points on this figure are obtained by fitting only the exponential piece of the escape time distribution to obtain the rate constant.}
    \label{fig:tau_scaling}
\end{figure}

In Figure \ref{fig:tau_scaling}, the late-time decay timescale of the wait time distribution is shown as a function of $\tau_0$. It is shown that the time constant of exponential decay in fitted Monte Carlo escape time distributions converges with $\gamma_{11}^{-1}$ at sufficiently high $\tau_0$, following a $t\propto(a\tau_0)^{1/3}$ scaling. The coefficient of this scaling ($0.51$) is within a factor of 2 of the approximate ``light-trapping time'' defined in \citet{2020MNRAS.497.3925L}, which predicts $ct/R=0.901(a\tau_0)^{1/3}$. At lower $\tau_0$, the effects of line core scattering are most important, leading to a larger discrepancy in the characteristic escape timescale. Here, the Monte Carlo accurately includes the photons which scatter in the core many times before escaping, while the semi-analytic solution does not capture this behavior as it uses only the Lorentzian piece of the line profile, and also does not use enough spatial modes to accurately model the frequency regime near line center. However, as $\tau_0$ grows, the effect of core scattering becomes smaller and the approximations hold, agreeing better with the expected $(a\tau_0)^{1/3}$ scaling \citep{1975ApJ...201..350A} for the rate constant at late times. The excess in the Monte Carlo data points due to core scattering decreases exponentially at higher $\tau_0$, and though these points are not shown at $\tau_0=10^8$ and $10^9$ due to computational expense, it is expected that the fractional error between the Monte Carlo and the $(a\tau_0)^{1/3}$ scaling would be less than 2\% at $\tau_0=10^9$.

We now evaluate the time-integrated spectrum (fluence) of the response to an impulse and compare it with the solution for the $H_0$ steady-state spectrum (Equation \ref{eq:H0surf}). Integrating Equation (\ref{eq:dEdtdnu}) over all times and dividing by the energy $E$, we find the fluence
\be \label{eq:spectrum}
P(x) & = &  \frac{16\pi^2 R \Delta}{3k\phi E}  \sum_{nm} (-1)^{n+1} \gamma_{nm}^{-1} J_{nm}(\sigma).
\ee
Integrating over $\nu$ then gives unity as required by the sum rule in Equation (\ref{eq:sumrule}). 

\begin{figure}
    \centering
    \includegraphics{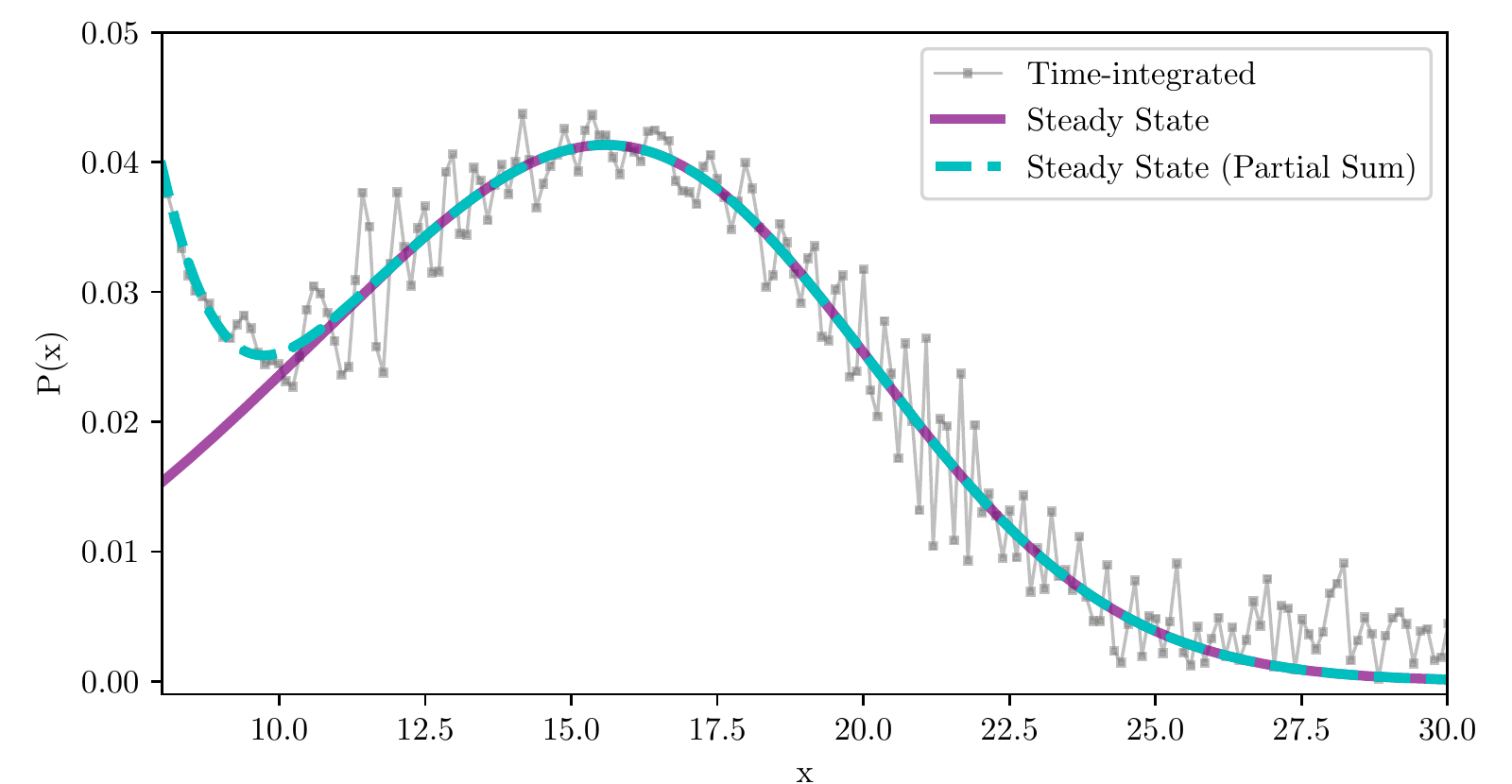}
    \caption{Fluence $P(x)$ vs. $x$ (see Equation \ref{eq:spectrum}). Fluence is the radiation flux integrated over time. Steady-state and time-integrated spectra for $n=1, ..., 20$ and $m=1, ..., 500$ are shown with $x_s=0$ and $\tau_0=10^7$. Note that the x-axis begins near the edge of the line core, as we are only concerned with the solutions' accuracy near the line wing.}
    \label{fig:steadystate}
\end{figure}

In Figure \ref{fig:steadystate}, the fluence for $x_s=0$ and $\tau_0=10^7$ is shown for a sum up to $n=20$ and $m=500$, labelled ``Time-integrated'', and is compared with two analytic solutions: the steady-state $H_0$ solution (Equation \ref{eq:H0surf}), labelled ``Steady State'', and the result for summing a finite number of spatial modes in the steady-state eigenfunction expansion as in the first line of Equation (\ref{eq:H0surf}), labelled ``Partial Sum''. Additional spatial modes $n$ increase the solutions' accuracy in the core of the line. If more spatial modes are included, the agreement with the steady-state spectrum extends further toward the line core. If additional frequency modes are included, faster-oscillating terms are incorporated into the Fourier sum over eigenmodes which create more perfect cancellations with the lower-order terms, reducing the ``ringing'' seen in the time-integrated spectrum. Extending the calculation deep into the line core by adding additional spatial modes could have an impact on the accuracy of the escape time distribution, but this would primarily affect the distribution at early times since the late time distribution is determined by the lowest order modes. This was the motivation for choosing a comparatively low number of spatial eigenmodes with respect to the number of frequency eigenmodes calculated.

\begin{figure}
    \centering
    \includegraphics{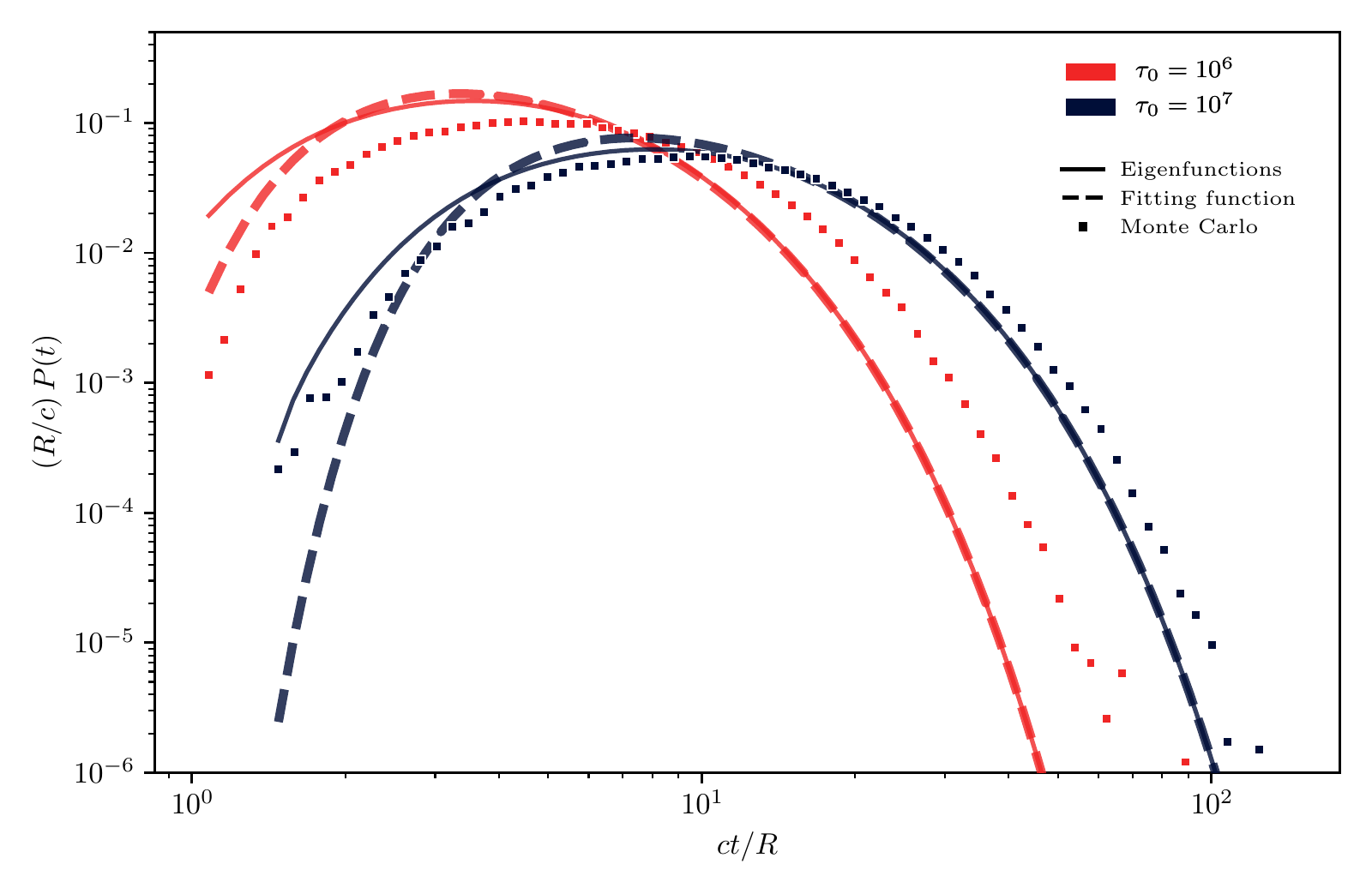}
    \caption{Wait-time distribution of escaping photons $P(t)$ vs. t  for $\tau_0=10^6$ and $10^7$, including the fitting function from Equation (\ref{eq:fitting_function}). A sum over 20 spatial eigenmodes and 500 frequency eigenmodes is labeled ``Eigenfunctions''. All calculations were performed with a monochromatic source of photons at line center ($x_s = 0$).}
    \label{fig:escape_time}
\end{figure}

In Figure \ref{fig:escape_time}, the escape time distributions calculated from Equation (\ref{eq:waittime}) are shown alongside Monte Carlo and the fitting function Equation (\ref{eq:fitting_function}) for $\tau_0=10^6, 10^7$ with $x_s=0$. The disagreement between the tail of the distribution and the Monte Carlo data is due to line core scattering which is not modeled by the eigenfunction solution, but improves for larger optical depth as seen in the figure. A large number of scatterings in the Doppler core affects the tail of the escape-time distribution, since photons with frequencies near line center will take longer to escape. Thus, the rate constant for the exponential falloff is overestimated slightly in the eigenfunction solution as compared with the Monte Carlo. The error in this rate constant is a function of $\tau_0$ since the effect from the Doppler core is greatest when it extends into the peak of the spectrum.

\section{Discussion}

\subsection{Steady-State Source}
A primary goal of this work is to present a solution for resonant scattering of photons near the line-center frequency $\nu_0$ in a uniform sphere. We have generalized a spherically symmetric solution derived by \citet{2006ApJ...649...14D} (called $H_0$ here) to allow a monochromatic source of photons with frequencies away from line center. We introduce a new term to this solution, $J_{\rm bc}$, which allows the boundary condition $J=\sqrt{3}H$ to be satisfied at the surface of the sphere. This is solved using a continuous Fourier expansion in frequency. The integrals are discretized and the Fourier coefficients solved for numerically. The resulting flux correction, $H_{\rm bc}$, scales as $H_0(a\tau_0)^{-1/3}$. Thus, for large $a\tau_0$, only a small correction to $H_0$ is needed, while larger errors are present in calculations performed at lower $a\tau_0$. Since the Laplacian form for frequency redistribution in the differential equation is only correct for photons in the wing where the line profile is $\phi \approx a/(\pi x^2 \Delta)$, our solutions do not accurately model the Doppler core of the \lya line. Because the peak of the spectral energy distribution of escaping photons is $x_{\rm peak} {\sim} (a\tau_0)^{1/3}$, calculations performed at small $a\tau_0$ are inaccurate due to the close proximity of the spectral peak and the Doppler core of the line.

By comparison with Monte Carlo simulations, we have shown that the enforcement of the correct frequency-dependent boundary conditions improves the accuracy of these analytic solutions for $a\tau_0 \gg 1$. Specifically, this solution shows improvement over previous solutions that utilized a $J=0$ surface boundary condition presented in \citet{1973MNRAS.162...43H}, \citet{1990ApJ...350..216N}, and \citet{2006ApJ...649...14D}. Several papers have previously compared these analytic models to Monte Carlo and seen discrepancies on the order of this correction. For example, in the top-left panel of Figure 1 from \citet{2006ApJ...649...14D}, the \lya spectrum emergent from a sphere of uniform optical depth is shown for $\tau_0=10^5, 10^6,$ and $10^7$ at a temperature of $T=10$ K, corresponding to $a=1.5 \times 10^{-2}$. The dotted line showing their theoretically-derived spectrum ($H_0$) displays an excess at the peak of at least 5-10 percent as compared with the Monte Carlo for $\tau_0=10^5$ and $10^6$. Another example is in \citet{2015MNRAS.449.4336S}, where the peak excess in the \lya spectrum is particularly noticeable for line center optical depths of $\tau_0=10^6$ and $10^7$ in the top panel of their Figure 5, which used slab geometry and a gas temperature of $T=10^4$ K. Again, the error in their solution is of order 5-10 percent. Both of these solutions are too large at the spectral peaks and too small further out in the wing, and the error scales approximately as $(a\tau_0)^{-1/3}$. We show in our Figure \ref{fig:sol_mc_tau} that the error present in $H_0$ is corrected by our treatment of the boundary condition at $\tau_0 = 10^7$ for T$=10^4$ K, corresponding to $a=4.72 \times 10^{-4}$. We note that our correction term $H_{\rm bc}$ is positive in the line wing and negative at the peak of the spectrum, which matches with the discrepancies noted in the aforementioned solutions.

\subsection{Impulsive Source}
The time-dependent transfer equation is solved in order to characterize the distribution of photon escape times. A semi-analytic approach is used, utilizing an expansion in space, time, and photon frequency. This boundary value problem in frequency $\sigma$ is solved to find the flux at the surface of the sphere as a function of $t$ and $\nu$. This solution is expressed as a sum over spatial and frequency modes $n$ and $m$, respectively. Calculating additional spatial eigenmodes increases the accuracy nearer to line center, but convergence is slow due to each eigenmode's weak dependence on $n$. Additional frequency eigenmodes introduce fast-oscillating terms that improve the accuracy of the Fourier sum, as their contributions cancel with components of lower-order terms to better represent the true solution. Integrating the solution over time produces a fluence that is shown to broadly agree with the steady-state calculations in Section \ref{sec:steadystate}, provided a sufficient number of terms in the sum and emission at the line-center frequency $\nu_0$. Integrating the solution over frequency leads to a distribution of photon escape time, which can be compared directly with Monte Carlo simulations. The sum over eigenmodes produces an escape-time distribution that broadly captures the behavior shown by Monte Carlo data---a rise at early times, transitioning to exponential decay in the tail of the distribution. It is expected that the accuracy of the rate constant for the tail of the distribution is limited by the effect of the Doppler core, which can trap photons at high optical depths until they diffuse outward in frequency, weighting the distribution toward later times. This physics is not modeled by our solution for two reasons: 1) our calculations ignored the Gaussian component of the Voigt line profile, leaving the Lorentzian piece which is accurate only in the line wing, and 2) knowing the core is not modeled accurately, we do not include a large enough number of spatial eigenmodes in the sum to resolve it. However, an approximate fitting function dependent on parameters $a$ and $\tau_0$ is found that adequately represents the escape time distribution of the Monte Carlo results within these constraints.

Our characterization of the escape time distribution leads to a possible application of this work. Models of the interaction of stellar \lya with the upper atmosphere of exoplanets and the associated transmission spectrum can be constructed with a treatment of resonant scattering in spherical geometry \citep{2017ApJ...851..150H, 2021ApJ...907L..47Y}. The Monte Carlo method can be used for this problem, but is limited by its high computational demand for large $\tau_0$ where there are many photon scatterings before escape. We seek to develop a method to accelerate the radiative transfer calculation. 

There are several methods that are commonly used to accelerate Monte Carlo radiation transfer calculations, including core skipping methods \citep{1968ApJ...153..783A,2002ApJ...567..922A} and hybrid diffusion methods \citep{2018MNRAS.479.2065S}. Another approach with wide application is modified random walk methods, such as those discussed in \citet{1984JCoPh..54..508F, 2009A&A...497..155M, 2010A&A...520A..70R}. In this approach, an outgoing photon is randomly sampled on the surface of the outgoing sphere by drawing its properties from distributions in outgoing frequencies, directions, and escape times, based on solutions to the diffusion equation. A method similar to this has been applied by \citet{2006ApJ...645..792T} to Lyman $\alpha$ transfer using the \citet{1990ApJ...350..216N} solution, but this solution of course does not utilize the frequency-dependent boundary condition at the surface of the sphere. Furthermore, to perform a full radiation hydrodynamic simulation with Monte Carlo acceleration, it will be necessary to calculate radiation forces within each cell due to \lya transfer. Similar calculations have been done in \citet{1976ApJ...208..286W} in plane-parallel geometry. However, these solutions are limited to optical depths below $2.5 \times 10^3$. For this work, it would be necessary to model line center optical depths of up to 1 million or more.

\section{Summary}

We have examined previous solutions to \lya transfer including resonant scattering in the limit of large optical depth, noting that the separation of variables and treatment of the boundary condition in \citet{1973MNRAS.162...43H}, \citet{1990ApJ...350..216N}, \citet{2006ApJ...649...14D} and others produces a discrepancy in the outgoing spectrum as compared with Monte Carlo. Here, we have derived the solution in spherical geometry with an appropriate treatment of the surface boundary condition. The key result is that the errors in the previously-cited works have been quantified via a correction term, $H_{\rm bc}$, which explains an excess in flux at the spectral peak and a deficit in the line wing of the calculated spectrum of \lya radiation as compared with Monte Carlo. The size of $H_{\rm bc}/H_0$ is of order unity when the spectral peaks are near the Doppler core, and diminishes at larger $\tau_0$ following a $(a\tau_0)^{-1/3}$ scaling. 

The time-dependent transfer equation for the impulsive source is solved numerically with an eigenfunction expansion. We demonstrate that it agrees with the steady-state spectrum for $x_s=0$ when integrated over time, though its rate of numerical convergence is slow and requires a sum over many modes to become accurate. The time-dependent solution is utilized to create wait-time distributions for photons escaping the sphere of optically-thick hydrogen gas. We compare the calculations from the time-dependent solution with Monte Carlo for a sample of $\tau_0$, noting general agreement in the resulting escape time distributions. The solution derived in our work here may be used as the basis for a novel implementation of the modified random walk method, which would accelerate Monte Carlo \lya transfer at large optical depths with potential applications in radiation hydrodynamic simulations of the atmospheres of exoplanets. 

\acknowledgments

This research was funded by NASA ATP grant 80NSSC18K0696, ``Exoplanetary MHD Outflows Driven by EUV Heating, Lyman alpha Radiation Forces and Stellar Tides". Resources supporting this work were provided by the NASA High-End Computing (HEC) Program through the NASA Advanced Supercomputing (NAS) Division at Ames Research Center. We thank the referee for a detailed report that helped improve the presentation of our work.
\restartappendixnumbering

\software{\texttt{numpy} \citep{2020NumPy-Array}, \texttt{scipy} \citep{2020SciPy-NMeth}, Coblis - Color Blindness Simulator (\href{https://www.color-blindness.com/coblis-color-blindness-simulator/}{color-blindness.com})}

\appendix
\section{ derivation of the transfer equation } \label{app:rteqn_derivation}

The problem is as follows. The radiative intensity $I = dE/(dA dt d\Omega d\nu)$ is the energy per perpendicular area $dA$, per time $dt$, per solid angle $d\Omega$ and per frequency $d\nu$ \citep{1986rpa..book.....R}. The intensity $I=I(\vec{x}, t, \vec{n}, \nu)$ will be considered a function of position $\vec{x}$, time $t$, photon (unit) direction vector $\vec{n}$, and cyclic frequency $\nu$. In the Eddington and two-stream approximations, $I(\vec{x},\nu) \simeq J(\vec{x},\nu) + 3 \vec{n} \cdot \vec{H}(\vec{x},\nu)$, where $J=(1/4\pi) \int d\Omega I$ is the mean intensity and $\vec{F} = 4\pi \vec{H}= \int d\Omega \vec{n} I$ is the flux.  

The transfer equation is \citep{1986rpa..book.....R}
\be
\frac{1}{c} \frac{\partial I}{\partial t} + \vec{n} \cdot \grad I & =& - \left( \alpha_{\rm sc} + \alpha_{\rm abs} \right) I + (1-p) j_{\rm sc} + j_{\rm em}.
\label{eq:rteqn}
\ee
The scattering coefficient, or inverse mean free path to scattering, is 
\be
\alpha_{\rm sc} & = & n_{\rm sc}\, \frac{\pi e^2}{m_e c}\, f\, \frac{\mathcal{H}(x,a)}{\sqrt{\pi} \Delta}
= k \phi.
\ee
The absorption coefficient $\alpha_{\rm abs}$, or inverse mean free path to true absorption, is a sum over species number density times absorption cross section. Once the incoming photon has promoted the electron to an excited state, the collisional de-excitation probability is $p$, and hence only a fraction $1-p$ of the excitations lead to re-emission of photons. 

\citet{1973MNRAS.162...43H} first showed that the transfer equation for the mean intensity $J$ will satisfy a Poisson equation involving space and frequency. In this section we will briefly review the derivation of this equation including photon destruction terms and an emission term.

 ``Hummer Case II-b"  \citep{1962MNRAS.125...21H} will be used for the redistribution function, for which the incoming photon is absorbed by the atom according to the natural broadening profile in the rest frame, then is re-emitted with a dipole phase function $g(\vec{n},\vec{n}^\prime)=(3/16\pi)(1+[\vec{n}\cdot \vec{n}^\prime]^2)$, which is appropriate for a 1s-2p transition \citep{1982qe}, and then is averaged over a Maxwell-Boltzmann distribution of speeds for the atom. The result can be written
\be
j_{\rm sc}(\vec{x},\vec{n},\nu) & = & k \int \frac{ d^3v}{ \pi^{3/2} v_{\rm th}^3} e^{-v^2/v_{\rm th}^2}\, 
\int d\Omega^\prime \int d\nu^\prime \,
g(\vec{n},\vec{n}^\prime) 
\nonumber \\ & \times & 
\delta \left( \nu - \nu^\prime - \nu_0 \vec{v} \cdot (\vec{n}-\vec{n}^\prime)/c \right)
\left( \frac{\Gamma/4\pi^2}{ \left(\nu^\prime - \nu_0 - \nu_0 \vec{v} \cdot \vec{n}^\prime/c \right)^2 + (\Gamma/4\pi)^2 } \right)  \,
I(\vec{x},\vec{n}^\prime,\nu^\prime)
\nonumber \\ & = & 4\pi k \int d\Omega^\prime \int d\nu^\prime R(\vec{n},\nu; \vec{n}^\prime,\nu^\prime) I(\vec{x},\vec{n}^\prime,\nu^\prime),
\ee
which defines the Case II-b redistribution function
\be
R(\vec{n},\nu; \vec{n}^\prime,\nu^\prime) & = & \frac{ g(\vec{n},\vec{n}^\prime) }{ 4\pi }
\int \frac{ d^3v}{ \pi^{3/2} v_{\rm th}^3} e^{-v^2/v_{\rm th}^2}\,
\delta \left( \nu - \nu^\prime - \nu_0 \vec{v} \cdot (\vec{n}-\vec{n}^\prime)/c \right)
\left( \frac{\Gamma/4\pi^2}{ \left(\nu^\prime - \nu_0 - \nu_0 \vec{v} \cdot \vec{n}^\prime/c \right)^2 + (\Gamma/4\pi)^2 } \right),
\ee
\citep{1952PASJ....4..100U, 1962MNRAS.125...21H}.

The integral of the redistribution function over outgoing and incoming frequency are
\be
\int d\nu\ R(\vec{n},\nu; \vec{n}^\prime,\nu^\prime) 
& = & \frac{1}{4\pi} g(\vec{n},\vec{n}^\prime) \phi(\nu^\prime)
\ee 
and
\be
\int d\nu^\prime \ R(\vec{n},\nu; \vec{n}^\prime,\nu^\prime) 
& = & \frac{1}{4\pi} g(\vec{n},\vec{n}^\prime) \phi(\nu)
\ee 
where the right hand side is the usual Voigt function, the thermal average of the Lorentzian. The former result implies that the integrated source and sink terms for scattering cancel for $p=0$. In addition, $4\pi R(\vec{n},\nu; \vec{n}^\prime,\nu^\prime)/\phi(\nu^\prime) $ is the normalized distribution for the outgoing $\vec{n}$ and $\nu$ given the incoming $\vec{n}^\prime$ and $\nu^\prime$. 

This probability distribution can be used to define the moments of the frequency shift \citep{1962ApJ...135..195O}
\be
\langle \delta \nu^n \rangle & = & \frac{ \int d\nu^\prime (\nu^\prime-\nu)^n R}{\int d\nu^\prime R}
= \frac{1}{\phi(\nu)}
\int \frac{ d^3v}{ \pi^{3/2} v_{\rm th}^3} e^{-v^2/v_{\rm th}^2}\,
\left( \frac{\nu_0 \vec{v} \cdot (\vec{n}^\prime-\vec{n}) }{c} \right)^n
\left( \frac{\Gamma/4\pi^2}{ \left(\nu - \nu_0 - \nu_0 \vec{v} \cdot \vec{n}/c \right)^2 + (\Gamma/4\pi)^2 } \right),
\ee
which are functions of $\nu$, $\vec{n}$ and $\vec{n}^\prime$. These integrals can be evaluated in terms of the dimensionless moments of the parallel velocity distribution, defined as
\be
\langle u_\parallel^n \rangle(x,a) & = & \frac{a/\pi }{\mathcal{H}(x,a)} \int 
\frac{du_\parallel u_\parallel^n e^{-u_\parallel^2}  }{(x-u_\parallel)^2 + a^2}.
\ee
The end results for the first and second moments are
\be
\langle \delta \nu \rangle & = & -\langle u_\parallel \rangle \left( 1 - \vec{n} \cdot \vec{n}^\prime \right) \Delta 
\\
\langle \delta \nu^2 \rangle & = & 
\left[ \langle u_\parallel^2 \rangle
\left( 1 - \vec{n} \cdot \vec{n}^\prime \right)^2
+ \frac{1}{2} \left( 1 - \left( \vec{n} \cdot \vec{n}^\prime\right)^2 \right) \right] \Delta^2 .
\ee

For small frequency shifts $\nu^\prime-\nu$, the incoming intensity may be expanded as 
\be
I(\vec{x},\vec{n}^\prime,\nu^\prime) & \simeq  &
I(\vec{x},\vec{n}^\prime,\nu) 
+ 
\frac{\partial I(\vec{x},\vec{n}^\prime,\nu)}{\partial \nu }(\nu^\prime-\nu)
+ \frac{1}{2} \frac{\partial^2 I(\vec{x},\vec{n}^\prime,\nu)}{\partial \nu^2} (\nu^\prime-\nu)^2 + ...
\ee 
and the Fokker-Planck expansion of $j_{\rm sc}$ is \citep{1994ApJ...427..603R}
\be
j_{\rm sc}(\vec{x},\vec{n},\nu) & \simeq  & 4\pi k \int d\Omega^\prime \int d\nu^\prime R(\vec{n},\nu; \vec{n}^\prime,\nu^\prime) 
\left[ 
I(\vec{x},\vec{n}^\prime,\nu) + \frac{\partial I(\vec{x},\vec{n}^\prime,\nu)}{\partial \nu }(\nu^\prime-\nu)
+ \frac{1}{2} \frac{\partial^2 I(\vec{x},\vec{n}^\prime,\nu)}{\partial \nu^2 }(\nu^\prime-\nu)^2
\right]
\nonumber \\ 
& =& 
k\phi(\nu) \int d\Omega^\prime g 
\left[ 
I(\vec{x},\vec{n}^\prime,\nu) 
+ \frac{\partial I(\vec{x},\vec{n}^\prime,\nu)}{\partial \nu }
\langle \delta \nu \rangle
+ \frac{1}{2} \frac{\partial^2 I(\vec{x},\vec{n}^\prime,\nu)}{\partial \nu^2 }
\langle \delta \nu^2 \rangle
\right].
\ee 
To perform the angular integrals, the Eddington approximation for the angular dependence is inserted with the following result
\be
j_{\rm sc} & = & k\phi J - k\phi \Delta \langle u_\parallel \rangle \left( \frac{\partial J}{\partial \nu} - \frac{6}{5} \vec{n} \cdot \frac{\partial \vec{H}}{\partial \nu} \right)
+ \frac{1}{2} \Delta^2 k\phi \left[ 
\frac{\partial^2 J}{\partial \nu^2} \left( \frac{7}{5} \langle u_\parallel^2 \rangle + \frac{3}{10} \right)
- \frac{12}{5} \langle u_\parallel^2 \rangle 
\vec{n} \cdot \frac{\partial^2 \vec{H}}{\partial \nu^2} \right]
\nonumber \\ & \simeq & 
k\phi J - k\phi \Delta \langle u_\parallel \rangle  \frac{\partial J}{\partial \nu} 
+ \frac{1}{2} \Delta^2 k\phi \left( \frac{7}{5} \langle u_\parallel^2 \rangle + \frac{3}{10} \right)
\frac{\partial^2 J}{\partial \nu^2} .
\label{eq:jsc}
\ee
The first term in Equation (\ref{eq:jsc}), $k\phi J$, represents re-emission of the photon through de-excitation of the atom. It cancels the $-k\phi J$ term in Equation (\ref{eq:rteqn}) that corresponds to excitation of the atom. The terms involving frequency derivatives of $\vec{H}$, if carried through the calculation, end up giving terms smaller than the largest terms by a factor of $1/x^2$, which is small in the line wing. These terms are ignored from here onward.

If only scattering is included, the transfer equation becomes
\be
\frac{1}{c} \frac{\partial }{\partial t} \left( J + 3\vec{n} \cdot \vec{H} \right)
+ \vec{n} \cdot \grad \left( J + 3\vec{n} \cdot \vec{H} \right)
& = & -3k\phi \vec{n} \cdot \vec{H} - k\phi \Delta \langle u_\parallel \rangle  \frac{\partial J}{\partial \nu} 
+ \frac{1}{2} \Delta^2 k\phi \left( \frac{7}{5} \langle u_\parallel^2 \rangle + \frac{3}{10} \right)
\frac{\partial^2 J}{\partial \nu^2}.
\ee
Integrating over angle and frequency then gives
\be
\frac{1}{c} \frac{\partial J(\vec{x}) }{\partial t} +  \grad \cdot \vec{H}(\vec{x}) & = & \int d\nu
\left( - k\phi \Delta \langle u_\parallel \rangle  \frac{\partial J}{\partial \nu} 
+ \frac{1}{2} \Delta^2 k\phi \left( \frac{7}{5} \langle u_\parallel^2 \rangle + \frac{3}{10} \right)
\frac{\partial^2 J}{\partial \nu^2}
\right) 
\nonumber \\ & =& 
k \int d\nu J \frac{\partial }{\partial \nu} 
\left( \phi \Delta \langle u_\parallel \rangle
+ \frac{\partial }{\partial \nu}\left[ \frac{1}{2} \phi \Delta^2 
\left( \frac{7}{5} \langle u_\parallel^2 \rangle + \frac{3}{10}  \right) \right]
\right),
\ee
where $\vec{H}(\vec{x})$ is the frequency integrated flux, and $\grad \cdot \vec{H}(\vec{x})  = 0 $ if there are no sources or sinks of radiation. Integration by parts has been used to factor $J$ out, assuming each term goes to zero at infinity. The quantity inside parentheses must be constant, and since each term should go to zero at infinity, that constant is zero. Hence the first and second moments of the frequency shift are related by
\be
\phi \Delta \langle u_\parallel \rangle
& = & -  \frac{\partial }{\partial \nu}\left[ \frac{1}{2} \phi \Delta^2 
\left( \frac{7}{5} \langle u_\parallel^2 \rangle + \frac{3}{10}  \right) 
\right].
\ee
As an example, in the damping wing, the line profile can be approximated as 
\be \label{eq:app:line_profile_wing}
\phi \simeq \frac{a}{\pi x^2 \Delta}
\ee
with $\langle u_\parallel \rangle \simeq 1/x$ and $\langle u_\parallel^2 \rangle \simeq 1/2$, and so this identity is satisfied. The scattering source function can then be rewritten
\be
j_{\rm sc} & \simeq & k\phi J + \frac{1}{2} k \Delta^2 \frac{\partial }{\partial \nu} 
\left[ \phi  \left( \frac{7}{5} \langle u_\parallel^2 \rangle + \frac{3}{10}  \right)\frac{\partial J}{\partial \nu}  \right]
\simeq k\phi J + \frac{1}{2} k \Delta^2 \frac{\partial }{\partial \nu} 
\left( \phi \frac{\partial J}{\partial \nu}  \right),
\ee
The following equations will use the approximations for the damping wing. Thus far the transfer equation is
\be
\frac{1}{c} \frac{\partial }{\partial t} \left( J + 3\vec{n} \cdot \vec{H} \right) + \vec{n} \cdot \grad \left( J + 3 \vec{n} \cdot \vec{H} \right)
& =& j_{\rm em}
- \left( k\phi + \alpha_{\rm abs} \right) \left( J + 3 \vec{n} \cdot \vec{H} \right)
+
(1-p) \left[
k\phi J + \frac{1}{2} k \Delta^2 \frac{\partial }{\partial \nu} 
\left( \phi \frac{\partial J}{\partial \nu}  \right)
\right]
\nonumber \\ & \simeq & 
j_{\rm em} - 3(k\phi + \alpha_{\rm abs}) \vec{n} \cdot \vec{H} - \left( p k\phi  + \alpha_{\rm abs} \right) J 
+ \frac{1}{2} k \Delta^2 \frac{\partial }{\partial \nu} 
\left( \phi \frac{\partial J}{\partial \nu}  \right),
\label{eq:rteqn2}
\ee
where leading order dissipative terms were kept in the second equality.
The moment equations are
\be
\frac{1}{c} \frac{\partial J}{\partial t} + \grad \cdot \vec{H} & = & j_{\rm em} 
- \left(p k\phi +  \alpha_{\rm abs} \right) J
+ \frac{1}{2} k \Delta^2 \frac{\partial }{\partial \nu} 
\left( \phi \frac{\partial J}{\partial \nu}  \right)
\ee
and
\be
\frac{1}{c} \frac{\partial \vec{H}}{\partial t} + \frac{1}{3} \grad J & =& - \left( k \phi + \alpha_{\rm abs} \right) \vec{H}.
\ee
The $\partial \vec{H}/\partial t$ term may be dropped for slowly changing sources.
Assuming the coefficients are constant in space, these two equations can be combined together to find
\be
\frac{1}{c} \frac{\partial J}{\partial t} - \frac{1}{3 (k\phi + \alpha_{\rm abs})} \nabla^2 J
& =& j_{\rm em} 
- \left( p k\phi +  \alpha_{\rm abs} \right) J
+ \frac{1}{2} k \Delta^2 \frac{\partial }{\partial \nu} 
\left( \phi \frac{\partial J}{\partial \nu}  \right).
\ee
Making the change of variables to $d\sigma$ using
\be \label{eq:change_of_variables}
d\sigma = \sqrt{\frac{2}{3}}\frac{d\nu}{\phi \Delta^2},
\ee
the equation can be rewritten in the standard form \citep{1973MNRAS.162...43H}
\be
-3 \left( \frac{k\phi + \alpha_{\rm abs}}{c}\right) \frac{\partial J}{\partial t} + \nabla^2 J + \left( \frac{k}{\Delta} \right)^2 \left( 1 + \frac{\alpha_{\rm abs}}{k\phi} \right)\frac{\partial^2 J}{\partial \sigma^2} & = & 
-3 \left( k\phi + \alpha_{\rm abs}\right) j_{\rm em}
+ 3 \left( k\phi + \alpha_{\rm abs}\right) \left( pk\phi + \alpha_{\rm abs}\right) J.
\label{eq:finaleqn}
\ee
    
For emission with luminosity $L$, with a delta function in space $\delta^3(\vec{x} - \vec{x}_s)$ at source position $\vec{x}_s$, a delta function in frequency $\delta(\nu-\nu_{\rm s})$ at source frequency $\nu_{\rm s}$, and isotropic in angles, the emission coefficient is
\be
j_{\rm em} & = & \frac{L}{4\pi} \delta^3(\vec{x} - \vec{x}_s) \delta(\nu-\nu_{\rm s}).
\label{eq:jem}
\ee
Multiplying by a factor $-3k\phi$, as appears in Equation (\ref{eq:finaleqn}), the delta function in $\nu$ becomes a delta function in $\sigma$ of the form
\be
-3k\phi j_{\rm em}  &= & - \frac{ \sqrt{6} kL}{4\pi \Delta^2} \delta^3(\vec{x} - \vec{x}_s) \delta (\sigma - \sigma_{\rm s}),
\label{eq:jem_v2}
\ee
where $\sigma_s = \sigma(\nu_s)$. 

\section{WKB Approximation} \label{app:wkb}

In this appendix, we will find a WKB solution for the differential equation Equation (\ref{eq:wkb_differential_eqn}) subject to the boundary conditions in Equations (\ref{eq:matching_condition_1}) and (\ref{eq:matching_condition_2}).
We will find this solution for the case $\sigma_s = 0$, assuming $\sigma \geq 0$ for simplicity since the solution is symmetric about $0$.

Near the line center, there are two independent solutions which can be written
\be \label{eq:wkb_J}
J(n,\sigma,s) & = & a_1 \left( \frac{\sigma}{\sigma_{\rm tp}} \right)^{1/2} J_{-3/4}\left( k_c \left[ \frac{\sigma}{\sigma_{\rm tp} }\right]^{2/3} \right)
\nonumber \\ &  + &  a_2 \left( \frac{\sigma}{\sigma_{\rm tp}} \right)^{1/2} J_{3/4}\left( k_c \left[ \frac{\sigma}{\sigma_{\rm tp} }\right]^{2/3} \right),
\ee
where $J_\alpha(x)$ are Bessel functions of the first kind, $a_1$ and $a_2$ are normalization constants, $\sigma_{tp}$ is given in Equation (\ref{eq:sigma_tp}), and $k_c=3\kappa_n \Delta \sigma_{\rm tp}/2k$. The solution far from line center is 
\be \label{eq:wkb_J_far}
J(n,\sigma,s) & = & a_3 \left( \frac{\sigma}{\sigma_{\rm tp}} \right)^{1/6} Ai\left(k_c^{2/3} \left[ \left( \frac{\sigma}{\sigma_{\rm tp}} \right)^{2/3} - 1 \right]\right),
\ee
where $Ai(x)$ is the Airy function of the first kind. Again, $a_3$ is a normalization constant. We can enforce the discontinuity in Equation (\ref{eq:matching_condition_2}) to find 
\be \label{eq:a2}
a_2 & = & - \frac{\Gamma(7/4) \sigma_{\rm tp}}{ 2^{1/4} k_c^{3/4}} \frac{6^{1/2}}{8} n^2 \frac{E}{kR^3}.
\ee
The discontinuity sets the value of $a_2$, and the corresponding term in Equation (\ref{eq:wkb_J}) is non-resonant. The discontinuity does not set the value of $a_1$ as that solution does not have a term linear in $\sigma$. Matched asymptotic expansions \citep{1999amms.book.....B} allows us to relate the normalization coefficients $a_1$ and $a_3$. The matching conditions are
\be \label{eq:a1/a2}
\frac{a_1}{a_2} & = & \frac{ \sin \left( \frac{\kappa_n \sigma_{\rm tp} \Delta}{k} + \frac{\pi}{8} \right) }{ \sin \left(  \frac{\kappa_n \sigma_{\rm tp} \Delta}{k} - \frac{\pi}{8} \right) }
\ee
\be \label{eq:a3/a2}
\frac{a_3}{a_2} & = & \frac{ k_c^{-1/3} }{ \sin \left(  \frac{\kappa_n \sigma_{\rm tp} \Delta}{k} - \frac{\pi}{8} \right) }.
\ee
By setting the denominator of this expression equal to zero and solving for the eigenfrequency $\gamma_{nm}$ contained in $\sigma_{\rm tp}$, we find the dispersion relation in Equation (\ref{eq:gamma_nm}).

While the solution above is specific to the case $\sigma_s = 0$, we can extend this approach to understand the case where $\sigma_s \neq 0$. In the interval $\sigma \in (0, \sigma_s)$, both the $Ai$ and $Bi$ Airy function solutions must be included, while for $\sigma \in (-\infty, 0)$ only $Ai$ is finite. The $Bi$ term causes the asymmetry. As $\gamma \rightarrow \gamma_{nm}$, however, it is small compared to the $Ai$ term, and hence the eigenfunctions are symmetric. 

\bibliography{ref.bib}{}
\bibliographystyle{aasjournal}

\end{document}